\documentclass[journal]{IEEEtran}

\usepackage{graphicx}
\usepackage{booktabs} % for \toprule/\midrule/\bottomrule
\usepackage[table]{xcolor} % table coloring

\usepackage{amsmath,amsfonts}
\usepackage{algorithmic}
\usepackage{array}
\usepackage{textcomp}
\usepackage{stfloats}
\usepackage{url}
\usepackage{verbatim}
\def\BibTeX{{\rm B\kern-.05em{\sc i\kern-.025em b}\kern-.08em
		T\kern-.1667em\lower.7ex\hbox{E}\kern-.125emX}}
\usepackage{balance}

\usepackage{multirow}
\usepackage{adjustbox} 
\usepackage{makecell}
\usepackage{arydshln}
\usepackage{subcaption}
\graphicspath{{figures/}}
\usepackage{algorithm}
\begin{document}
\title{Inferring Network Evolutionary History via Structure--State Coupled Learning}
\author{En~Xu,
	Shihe~Zhou,
	Huandong~Wang,
	Jingtao~Ding,
	and~Yong~Li
	\IEEEcompsocitemizethanks{\IEEEcompsocthanksitem En Xu, Shihe~Zhou, Huandong~Wang, Jingtao Ding, and Yong Li are with the Department of Electronic Engineering, Tsinghua University, Beijing, China. 
		(E-mail: xuen@mail.tsinghua.edu.cn, 
		zhoush23@mails.tsinghua.edu.cn,
		wanghuandong@tsinghua.edu.cn,
		dingjt15@tsinghua.org.cn,
		liyong07@tsinghua.edu.cn.)
	}% <-this % stops an unwanted space

}

\markboth{Inferring Network Evolutionary History via Structure--State Coupled Learning}%
{Xu \MakeLowercase{\textit{et al.}}: Inferring Network Evolutionary History via Structure--State Coupled Learning}

\maketitle

\begin{abstract}
Inferring a network's evolutionary history from a single final snapshot with limited temporal annotations is fundamental yet challenging. Existing approaches predominantly rely on topology alone, which often provides insufficient and noisy cues. This paper leverages network steady-state dynamics---converged node states under a given dynamical process---as an additional and widely accessible observation for network evolution history inference. We propose CS$^2$, which explicitly models structure--state coupling to capture how topology modulates steady states and how the two signals jointly improve edge discrimination for formation-order recovery. Experiments on six real temporal networks, evaluated under multiple dynamical processes, show that CS$^2$ consistently outperforms strong baselines, improving pairwise edge precedence accuracy by 4.0\% on average and global ordering consistency (Spearman-$\rho$) by 7.7\% on average. CS$^2$ also more faithfully recovers macroscopic evolution trajectories such as clustering formation, degree heterogeneity, and hub growth. Moreover, a steady-state-only variant remains competitive when reliable topology is limited, highlighting steady states as an independent signal for evolution inference.
\end{abstract}

\begin{IEEEkeywords}
Complex network, Network evolution, Network reconstruction, Steady-state dynamics.
\end{IEEEkeywords}

% !TeX root = TKDE-main.tex
\section{Introduction}
\label{sec:intro}

\begin{figure*}[t]
	\centering
	\includegraphics[width=0.95\textwidth]{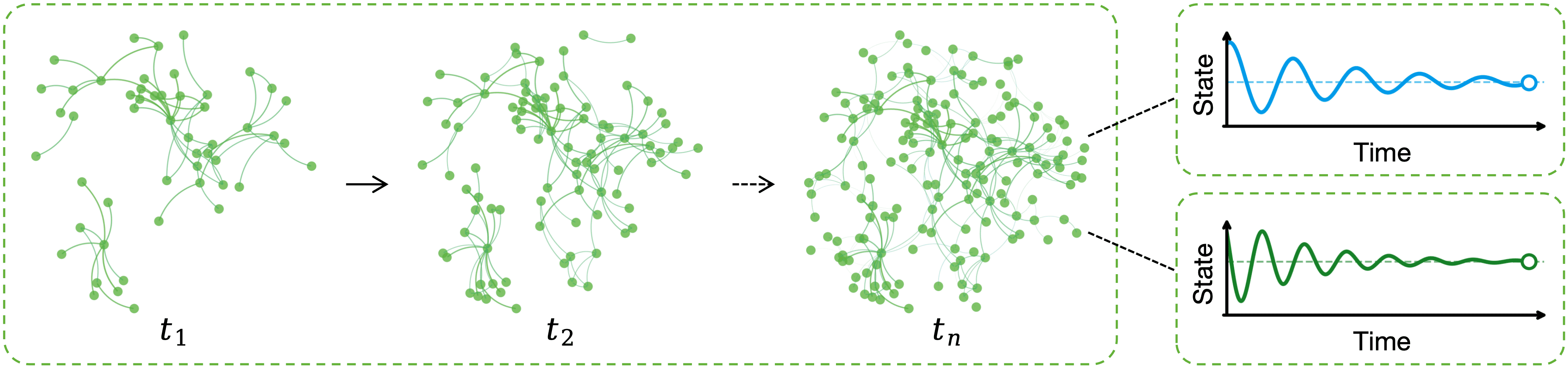}
	% \vspace{-0.3cm}
	\caption{Problem illustration. A temporal network evolves through sequential edge formation, but is often observed only as a final snapshot with scarce temporal annotations. The goal is to recover the edge formation order from the snapshot, optionally augmented with steady-state node observations.}
	% \vspace{-0.5cm}
	\label{fig:evolutionary_schematic}
\end{figure*}
%==================== 段 1 ====================%
%\paragraph*{P1：问题背景与重要性（What \& Why）}

In complex systems such as social networks~\cite{barabasi2002evolution}, urban systems~\cite{wang2023pattern}, and ecological networks~\cite{seferbekova2023spatial, ding2024artificial, xu2025survey}, network structure evolves continuously, whereas observations are often limited to a single snapshot at a target time (Figure~\ref{fig:evolutionary_schematic}). This snapshot is the end result of an unobserved sequence of local formation events, and it implicitly encodes the network's evolutionary history, key interactions, and generative mechanisms. Inferring this history from a final snapshot can therefore advance mechanistic understanding and support downstream analyses such as identifying sources of influence, explaining functional organization, and disentangling causal dependencies~\cite{liao2017ranking}. However, the final topology compresses the entire history into a single static form, making history inference from such a snapshot inherently challenging.

%==================== 段 2 ====================%
%\paragraph*{P2：现有研究与局限（What has been done \& Why insufficient）}
Recent work has begun to address this problem from the final topology alone. \textit{StructEvo}~\cite{wang2024reconstructing} learns a structure--time mapping from a small set of timestamped edges and ranks the remaining edges within the same network. \textit{TopoDiff}~\cite{Xu2026TKDE} extends this paradigm to unseen networks by training on multiple temporal networks to improve transferability, but it does not show a clear accuracy gain on the original within-network recovery task. More broadly, topology-only approaches assume complete and reliable structure at inference time; in practice, network observations can be noisy, incomplete, or even unavailable, which limits robustness and motivates complementary signals beyond topology.

%==================== 段 3 ====================%
%\paragraph*{P3：提出“利用稳态反推演化”的核心想法（New angle / key intuition）}
Beyond topology, we consider \emph{steady-state dynamics} as a complementary observation for evolution reconstruction~\cite{boccaletti2006complex}. A steady state is the converged node-state configuration under a given dynamical process, such as the final opinion profile~\cite{bu2019graph}, equilibrium diffusion concentration~\cite{cao2024survey}, or steady traffic flow~\cite{gallotti2015multilayer}. As a dynamical response to the underlying structure, the steady state is jointly shaped by topology and dynamics, and can therefore capture non-local dependencies that are not explicit in the final snapshot~\cite{millan2025topology}.
Steady states are informative for three reasons. (i) They provide a global projection of network interactions and can be sensitive to high-order structural patterns beyond local indices~\cite{zhu2018high}. (ii) Different edge formation orders induce different trajectories toward convergence, leaving accumulated traces in the observed steady state that correlate with formation times~\cite{panahi2023rate}. (iii) In many systems, edge formation is state-driven (e.g., homophily in social networks~\cite{xu2024temporal} and expression-dependent regulation in biological networks~\cite{jin2013evolutionary}), creating an intrinsic coupling between steady states and evolution order.
Steady states are also practical: they are often available from logs, monitoring, or aggregated statistics~\cite{liu2025beyond}, even when the underlying interaction network is noisy, incomplete, or unobservable~\cite{li2019fates}. This motivates leveraging steady states to improve temporal recovery and to enable inference under structure-limited conditions.

%==================== 段 4 ====================%
%\paragraph*{P4：本文提出的任务与方法框架（What we propose）}
In this paper, we study how to recover the complete edge formation order when only partial edge timestamps are available. Our key idea is to augment the final snapshot with its associated steady state and to learn a \emph{structure--state coupling} representation for temporal ordering, so that the model can exploit both topological cues and the dynamical response induced by the same structure.
We propose Coupled Structure--State (CS$^2$), a dynamics-informed learning framework that performs temporal ordering through pairwise edge comparisons and then reconstructs a global sequence via Borda aggregation~\cite{emerson2013original}. CS$^2$ extracts structural and steady-state features for each candidate edge and explicitly couples the two modalities to improve edge discrimination for ordering.
Experiments demonstrate that coupling structure and steady state yields consistent gains for network evolution history inference. Importantly, when reliable topology is incomplete or costly to obtain, steady-state signals alone remain competitive, indicating that steady states provide an independent source of time-discriminative cues and enabling practical inference under structure-limited scenarios.

%==================== 段 5 ====================%
%\paragraph*{P5：贡献总结（Contributions）}
Our main contributions are summarized as follows:
\begin{itemize}
	\item We introduce steady-state dynamics as a complementary and widely accessible signal for network evolution history inference, and provide a principled justification of why steady states can carry time-discriminative information. We further show that steady-state signals alone can remain competitive when reliable topology is incomplete or unavailable.
	
	\item We propose CS$^2$, a structure--state coupled learning framework that explicitly models interactions between topology and steady states to improve edge discrimination for temporal ordering.
	
	\item We show that CS$^2$ achieves consistent improvements on both pairwise and global order recovery, improving pairwise edge precedence accuracy by 4.0\% and global ordering consistency (Spearman-$\rho$) by 7.7\% on average over the strongest baselines, while also better recovering macroscopic evolution trajectories such as clustering formation, degree heterogeneity, and hub growth.
\end{itemize}

% !TeX root = TKDE-main.tex
\section{RELATED WORK} \label{sec:Relatedwork}

\subsection{Temporal Network}
Research on temporal networks spans multiple settings, but most work is not tailored to the \emph{inverse} problem studied here---recovering the formation order of existing edges from a single final snapshot with scarce temporal annotations.

A classic line models network evolution through generative mechanisms, prescribing growth rules for nodes and edges. Representative examples include preferential attachment (Barabási--Albert)~\cite{barabasi1999emergence}, small-world rewiring (Watts--Strogatz)~\cite{watts1998collective}, the PSO model~\cite{papadopoulos2012popularity}, and fitness-based models~\cite{bianconi2001competition}. These models explain how macroscopic statistics (e.g., degree distribution, clustering, community structure) can emerge, but they are designed for simulation and synthesis rather than for recovering the realized edge formation order from an observed network.

Another major line develops deep learning methods for fully observed temporal interaction streams, such as TGAT~\cite{tgat_iclr20}, TGN~\cite{tgn_icml_grl2020}, DySAT~\cite{sankar2020dysat}, and EvolveGCN~\cite{egcn}. These methods assume a time-stamped stream or a sequence of snapshots as input and learn evolving representations via temporal attention, time encoding, recurrent updates, or message passing, primarily to forecast future structure (e.g., upcoming edges) or states. In contrast, our setting provides only a single final snapshot with missing timestamps; the goal is to infer the \emph{historical} ordering of existing edges rather than to predict future events. This fundamental difference in both observability and objective makes temporal-stream models not directly applicable to evolution reconstruction from a static snapshot~\cite{holme2012temporal,millidge2024predictive}.

Closest to our setting are learning-based evolution reconstruction methods that infer edge order under partial timestamp supervision. \textit{StructEvo}~\cite{wang2024reconstructing} (Nature Communications) learns a structure--time mapping from a small set of labeled edges and ranks the remaining edges within the same network using topological patterns in the final snapshot. \textit{TopoDiff}~\cite{Xu2026TKDE} extends this paradigm to unseen networks by training on multiple temporal networks with synthetic temporal augmentation, improving cross-network transferability.

Despite this progress, topology-only reconstruction remains limited in practice. A final snapshot encodes connectivity but not the dynamical responses or state-dependent mechanisms that often accompany edge formation, which can reduce identifiability in complex systems. In addition, accurate topology can be costly to obtain and may be noisy or incomplete, making purely topology-driven inference less reliable and motivating complementary observational signals beyond structure.

\subsection{Steady-State Dynamics on Networks}
Network dynamics provides a well-established lens for understanding how information, opinions, or states evolve on graphs. Typical processes include diffusion~\cite{cao2024survey}, random walks~\cite{masuda2017random}, synchronization~\cite{vasseur2014synchronous}, opinion dynamics~\cite{bu2019graph}, and epidemic spreading~\cite{ma2022patient}. Under a fixed topology, such dynamics often converge to a steady state, which is jointly determined by the graph structure and the dynamical rule. Because it aggregates multi-step interactions over the entire network, the steady state can be interpreted as a global response of topology, complementary to local structural indices.

Prior studies have shown that steady states and related diffusion-equilibrium quantities reflect high-order structural properties, including community organization~\cite{wen2021gravity}, centrality heterogeneity~\cite{evans2022linking}, and global connectivity patterns~\cite{ji2021survey}. This perspective is also widely used in representation learning and spectral graph theory: heat kernel embeddings~\cite{saito2022hypergraph}, diffusion state distance~\cite{liu2022stationary}, random-walk/diffusion-based embeddings~\cite{xing2023diff}, and Laplacian spectral decompositions~\cite{gallagher2021spectral} exploit convergence behaviors to capture global similarities and latent structural patterns that are difficult to recover from one-shot topology descriptors.

Despite their broad use for structure characterization and embedding learning, steady states have rarely been explored as an observational signal for \emph{temporal} inference on static snapshots. In particular, how to leverage steady-state observations to recover edge formation order and reconstruct network evolution history remains largely open. This gap motivates the steady-state-driven evolution reconstruction framework proposed in this paper.

% !TeX root = TKDE-main.tex
\section{Preliminaries} \label{sec:pre}

\subsection{Reconstructing Network Evolution via Pairwise Comparisons}

Directly predicting an absolute formation time for each edge is often unnecessary and unstable for evolution reconstruction. Edge times are continuous and typically comparable only within the same network, while the relative order is the core object of interest. Under partial temporal supervision, time regression can therefore be sensitive to noise and label sparsity.

We instead predict \emph{pairwise temporal precedence}: given two edges, decide which one formed earlier. This converts the original regression problem into a binary classification problem that is easier to optimize and more robust under scarce labels. Once pairwise relations are available, they can be aggregated to produce a complete ordering over all edges.

We further establish a quantitative link between local pairwise accuracy and global ordering quality. Let \(p\) denote the probability of correctly predicting precedence for a randomly sampled edge pair. Then, the expected error of the global ordering recovered from pairwise outcomes satisfies
\begin{equation}
	\mathcal{E}_\text{theory} =
	\sqrt{\frac{p(1-p)}{(2p-1)^2}}
	\frac{1}{\sqrt{M}},
	\label{eq:equivalence}
\end{equation}
where \(M\) is the number of edges. The full derivation is provided in Appendix~\ref{apdx:equal}.

% Eq.~\eqref{eq:equivalence} implies two key observations: (i) global ordering error decreases with \(M\); and (ii) a reasonably accurate global sequence can be recovered without requiring near-perfect pairwise accuracy. This supports pairwise prediction as a stable surrogate objective for edge-order recovery.

After obtaining pairwise precedence predictions, we reconstruct the global ordering using the Borda count. It assigns each edge a score equal to the number of pairwise ``wins'' and returns the ordering most consistent with the predicted comparisons. Details are given in Algorithm~\ref{alg:borda_reconstruction}.

\begin{figure}[t]
	\centering
	\includegraphics[width=0.4\textwidth]{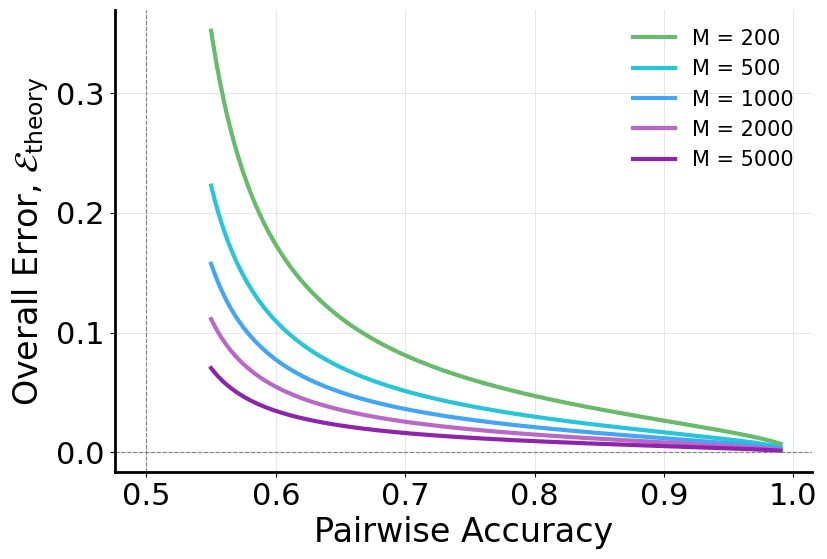}
	% \vspace{-0.3cm}
	\caption{Theoretical link between pairwise precedence accuracy and the expected error of the recovered global ordering (Eq.~\eqref{eq:equivalence}).}
	% \vspace{-0.5cm}
	\label{fig:rmse_acc}
\end{figure}

In summary, pairwise learning aligns better with the objective of recovering the edge formation order and offers an optimization-friendly route to reconstructing network evolution. Figure~\ref{fig:rmse_acc} visualizes the theoretical connection between pairwise accuracy and global ordering error implied by Eq.~\eqref{eq:equivalence}.

\subsection{Problem Definition}

We consider an undirected network fully observed at a target time,
\[
G=(V,E),\qquad |V|=N,\ |E|=M,
\]
where edges in \(E\) are generated sequentially by an unknown process. For each edge \(e\in E\), we denote its ground-truth formation order by a normalized time
\[
\alpha_e \in [0,1],
\]
where smaller values indicate earlier formation.

Among all edges, only a subset has observable formation times, denoted as the labeled set
\[
E_{\mathrm{lab}}=\{e_i \mid \alpha_{e_i} \ \text{known}\},
\]
and the remaining edges are unlabeled,
\[
E_{\mathrm{unk}} = E \setminus E_{\mathrm{lab}}.
\]

In addition to topology, we observe a steady-state vector obtained by running a dynamical process (e.g., diffusion, gene regulation, or opinion dynamics) on \(G\) until convergence,
\[
\mathbf{x} = (x_1, x_2, \ldots, x_N),
\]
where \(x_i\) is the steady-state value of node \(i\). The steady state is jointly determined by the structure \(G\) and the dynamical rule, and can be viewed as a global response of topology under the dynamics.

Given \((G,\mathbf{x})\) and partial temporal labels \(E_{\mathrm{lab}}\), our goal is to reconstruct the complete edge formation order. Specifically, we define:

1.  \textit{Pairwise ground-truth precedence.}
For any two edges \(e_i, e_j \in E\), define their ground-truth precedence relation as
\[
R(e_i,e_j)=
\begin{cases}
	1, & \alpha_{e_i} < \alpha_{e_j},\\[2mm]
	0, & \alpha_{e_i} \ge \alpha_{e_j}.
\end{cases}
\]
Here, \(R(e_i,e_j)=1\) indicates that \(e_i\) forms earlier than \(e_j\).

2. \textit{Pairwise order prediction.}
We learn a decision function
\[
\widehat{R}(e_i,e_j)=f\big(G,\mathbf{x},e_i,e_j\big),
\]
to predict the precedence for any edge pair.

3. \textit{Global order reconstruction.}
From the pairwise predictions \(\widehat{R}(e_i,e_j)\), we apply a rank aggregation method (Borda count in this paper) to construct a global ordering that is maximally consistent with local comparisons,
\[
\hat{\pi}: E \rightarrow \{1,2,\ldots,M\},
\]
where \(\hat{\pi}(e)\) denotes the position of edge \(e\) in the reconstructed sequence.

The predicted evolution sequence can be written as
\[
\hat{T} = \{e_{k_1}, e_{k_2}, \ldots, e_{k_M}\},
\]
where \(k_1,k_2,\ldots,k_M\) are edge indices sorted by the predicted order.

Our objective is to maximize the ranking consistency between the reconstructed sequence \(\hat{T}\) and the ground-truth sequence \(T\). This formulation highlights the available observations (topology and steady state) and the desired output (a complete edge formation order), and serves as the foundation for the proposed model.

\subsection{Steady-States for Edge-Order Recovery}

Topology-only reconstruction relies solely on the final snapshot, whose observable cues can be weak or ambiguous. We therefore introduce the steady state \(\mathbf{x}\) associated with the observed network as an auxiliary signal under the same supervision. The usefulness of steady states for edge-order recovery can be understood from three complementary perspectives.

\textbf{(i) Global, structure-filtered response.}
Under diffusion, synchronization, random walks, or opinion dynamics, the steady state can be viewed as a structure-filtered global response of the network. It aggregates multi-hop interactions and can reflect high-order properties such as community organization, path bottlenecks, and centrality heterogeneity, which are difficult to capture with local indices or one-shot topological descriptors. Hence, steady states can amplify subtle structural differences in node-state space (see Appendix~\ref{apdx:steady_highorder}).

\textbf{(ii) Sensitivity to formation history.}
Because edges appear sequentially, early edges affect the system for longer and can shape dynamical trajectories more strongly than late edges. Consequently, different formation orders can lead to different observed steady states even if the final topology is the same, leaving accumulated temporal traces that correlate with edge ``earliness'' (see Appendix~\ref{apdx:steady_pathdependence}).

\textbf{(iii) State-driven coevolution.}
In many adaptive and coevolving systems, node states influence how links form or rewire. For example, discordant links in coevolving opinion models are rewired toward like-minded nodes~\cite{holme2006nonequilibrium}, and susceptible individuals in adaptive epidemic networks avoid infected ones via link deletion and rewiring~\cite{gross2006epidemic}. Such state$\rightarrow$structure coupling implies that steady states are not incidental by-products, but are intertwined with the formation process and can thus carry discriminative cues about edge ordering.

In summary, steady states provide a global response of structure, retain traces of the formation history, and may be intrinsically coupled with the evolution mechanism. These properties complement topology and improve identifiability for edge-order recovery, especially when structural observations are noisy or incomplete.

% !TeX root = TKDE-main.tex
\section{Method} \label{sec:methods}

\subsection{Overview of the Framework}

\begin{figure*}[t]
	\centering
	\includegraphics[width=\textwidth]{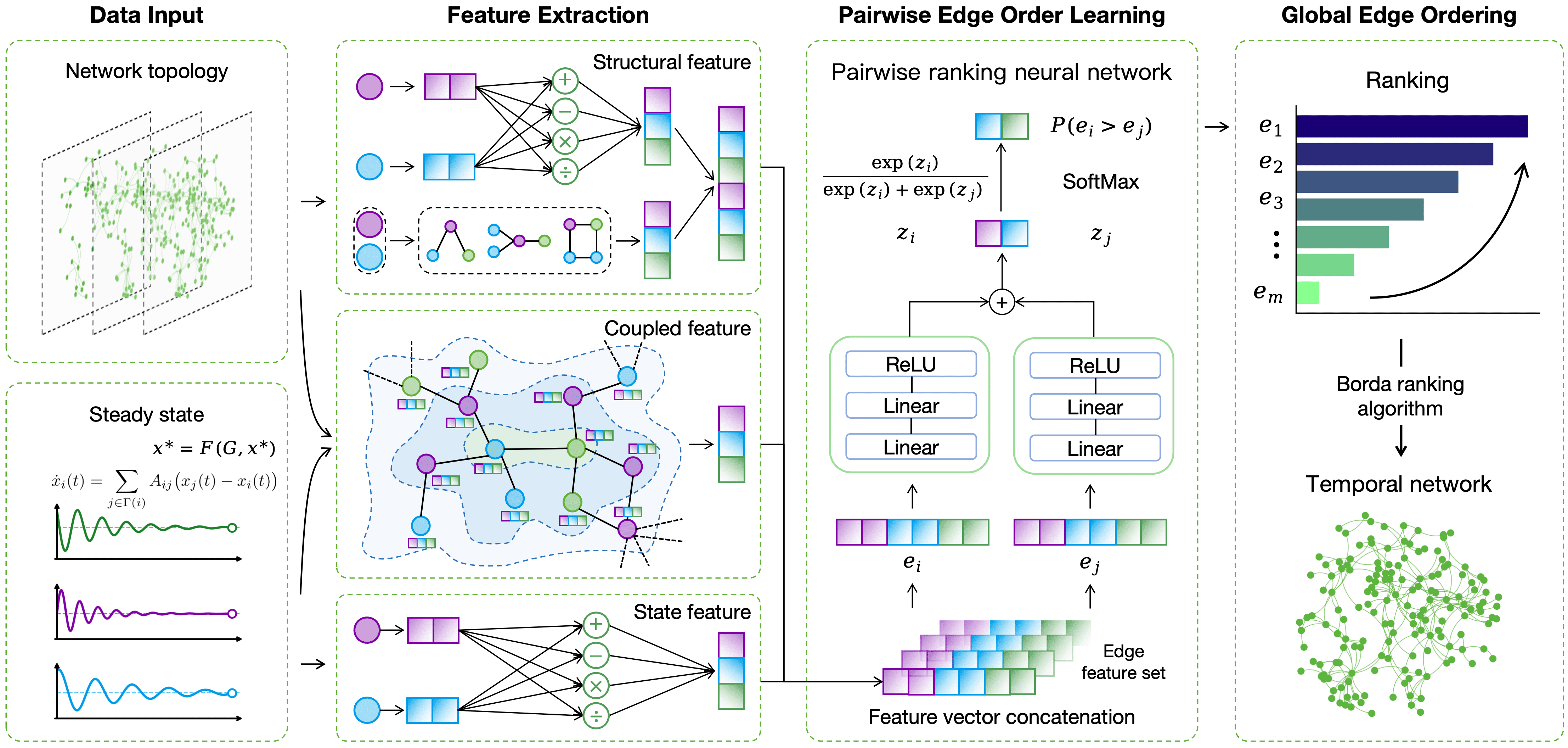}
	% \vspace{-0.3cm}
	\caption{Overview of CS$^2$. Given the final snapshot $G_T$ and its steady state $\mathbf{x}_T$, the framework extracts structural and steady-state features, learns coupled representations via graph propagation, predicts pairwise edge precedence with CPNN, and aggregates comparisons to recover a global edge formation order.}
	% \vspace{-0.5cm}
	\label{fig:framework}
\end{figure*}

We present CS$^2$, a structure--state coupled framework for edge-order inference. As illustrated in Figure~\ref{fig:framework}, CS$^2$ takes as input the final snapshot $G_T=(V,E_T)$ and its associated steady state $\mathbf{x}_T$, together with partial timestamp labels on a subset of edges. The output is a globally consistent ordering over all edges in $E_T$. The framework consists of four stages:

\textit{(1) Structure and steady-state feature extraction.}
From $G_T$, we compute structural node statistics and structural edge features. From $\mathbf{x}_T$, we construct steady-state edge features by composing endpoint states.
\textit{(2) Coupled edge feature construction.}
We feed node-level structure/state inputs into a graph propagation module to learn coupled node embeddings, and then transform them into edge-level coupled features that capture structure--state interactions beyond handcrafted descriptors.
\textit{(3) Pairwise temporal precedence prediction.}
We form a unified edge representation by concatenating structural, steady-state, and coupled features. CPNN maps each edge to a score and predicts the precedence for any edge pair by comparing their scores.
\textit{(4) Global order reconstruction.}
We aggregate all pairwise predictions with Borda count to produce a globally consistent edge sequence.

\subsection{Structure and Steady-State Feature Extraction}

As shown in the feature-extraction block of Figure~\ref{fig:framework}, we construct two types of edge features directly from the final observation $(G_T,\mathbf{x}_T)$: (i) structural edge features $\mathbf{f}^{\mathrm{struct}}(e)$ computed from topology; and (ii) steady-state edge features $\mathbf{f}^{\mathrm{state}}(e)$ composed from endpoint states. We also compute node-level structural statistics $\text{struct}(i)$, which are used as inputs to the graph propagation module in Section~\ref{sec:coupled_feat}. In our implementation, $\text{struct}(i)=[k(i),\,C(i),\,\mathrm{core}(i)]$, whereas $\mathbf{f}^{\mathrm{struct}}(e)$ includes edge-level descriptors derived from the endpoints and their neighborhoods (e.g., degree-based features, common neighbors and overlap indices, path-based indices, and centrality-derived metrics). Together with $\mathbf{f}^{\mathrm{state}}(e)$, these features form the raw inputs for coupling and for the final edge representation used by CPNN.

\subsubsection{Structural Features}

For the final network $G_T=(V,E_T)$, let $A$ be the adjacency matrix. The neighbor set and degree of node $i$ are defined as
\[
\Gamma(i)=\{j\in V : A_{ij}=1\}, \qquad k(i)=|\Gamma(i)|.
\]
To avoid numerical instability in ratio and logarithm operations, we use a small constant $\varepsilon>0$ when needed. We extract multi-scale structural cues, including local connectivity, neighborhood overlap, short-path structures, and centrality-related signals.

\textit{Node-level features.}
We start from three basic node statistics: degree $k(i)$, clustering coefficient $C(i)$, and coreness $\mathrm{core}(i)$ from $k$-core decomposition. They provide a compact summary of local density and meso-scale membership, and are used both directly and for constructing edge-level descriptors.

\textit{Edge-level features.}
For each edge $e=(i,j)$, we derive descriptors from the endpoints and their neighborhoods.

We include degree-based features (with endpoint clustering coefficients):
\begin{equation}
	\left\{
	\begin{aligned}
		&k(i),\; k(j),\; k(i)+k(j),\; k(i)k(j),\;
		\min\{k(i),k(j)\},\\
		&\max\{k(i),k(j)\},\; C(i),\; C(j)
	\end{aligned}
	\right\}
	\label{eq:deg_features}
\end{equation}

We also report the preferential attachment score, which corresponds to the degree product $k(i)k(j)$ in Eq.~\eqref{eq:deg_features}.

We include neighborhood-overlap metrics~\cite{liben2007}:
Let
\[
\mathrm{CN}(i,j)=|\Gamma(i)\cap\Gamma(j)|,\qquad
U(i,j)=|\Gamma(i)\cup\Gamma(j)|.
\]
We include $\mathrm{CN}(i,j)$ as the common-neighbor feature, and define
\begin{align}
	\mathrm{Jaccard}(i,j) &= \frac{\mathrm{CN}(i,j)}{U(i,j)+\varepsilon},\\
	\mathrm{AA}(i,j) &= \sum_{z\in \Gamma(i)\cap \Gamma(j)} \frac{1}{\log(k(z)+\varepsilon)},\\
	\mathrm{RA}(i,j) &= \sum_{z\in \Gamma(i)\cap \Gamma(j)} \frac{1}{k(z)}.
\end{align}

We include edge strength as a normalized neighborhood overlap:
\begin{equation}
	\mathrm{ES}(e)=\frac{\mathrm{CN}(i,j)}{k(i)+k(j)-2-\mathrm{CN}(i,j)+\varepsilon}.
\end{equation}

We include edge betweenness and edge clustering~\cite{santoro2022onbra}:
\begin{align}
	\mathrm{BN}(e) &=
	\sum_{a,b\in V}
	\frac{\sigma(a,b\mid e)}{\sigma(a,b)},\\
	\mathrm{CC}_{\mathrm{edge}}(e) &=
	\frac{\mathrm{CN}(i,j)}
	{\max\{\min(k(i)-1,k(j)-1),\,1\}}.
\end{align}

We include a local path index~\cite{zhou2009}:
\begin{equation}
	\mathrm{LP}(e) = (A^2)_{ij} + \lambda(A^3)_{ij}.
	\label{eq:lp}
\end{equation}
where $(A^k)_{ij}$ is the number of length-$k$ paths from $i$ to $j$.

We include PageRank- and core-derived edge features~\cite{bianchini2005inside}:
\begin{equation}
	\begin{aligned}
		\mathrm{PR}(e)&=\max\{\mathrm{PR}(i),\,\mathrm{PR}(j)\},\\
		\mathrm{KS}(e)&=\min\{\mathrm{core}(i),\,\mathrm{core}(j)\}.
	\end{aligned}
\end{equation}

\subsubsection{Steady-State Features}

Steady-state features are constructed from the final steady-state vector $\mathbf{x}_T$, where $x_u$ denotes the steady-state value of node $u$. For each edge $e=(i,j)$, we compose endpoint states to capture similarity and disparity patterns:
\begin{equation}
	\left[
	x_i,\; x_j,\; x_i + x_j,\; |x_i - x_j|,\; x_i x_j,\;
	\frac{x_i}{x_j+\varepsilon},\;
	\frac{x_j}{x_i+\varepsilon}
	\right]
\end{equation}

This defines the edge-level steady-state feature vector $\mathbf{f}^{\mathrm{state}}(e)$.

\subsubsection{Normalization}

To make different feature scales comparable, we normalize all edge-level features before feeding them into the model. Let $X\in\mathbb{R}^{M\times d}$ be the edge feature matrix. For each feature dimension $X_k$, we apply min--max normalization followed by standardization:

\begin{equation}
	\begin{aligned}
		X^{\mathrm{min\text{-}max}}_{k}
		&= \frac{X_{k} - \min(X_{k})}
		{\max(X_{k}) - \min(X_{k}) + \varepsilon}, \\[4pt]
		\tilde{X}_{k}
		&= \frac{X^{\mathrm{min\text{-}max}}_{k} - \mu_{k}}{\sigma_{k}},
	\end{aligned}
	\label{eq:normalization}
\end{equation}

where $\varepsilon$ is a small positive constant to avoid division by zero, and $\mu_k$ and $\sigma_k$ denote the mean and standard deviation of the $k$-th feature. The normalized matrix $\tilde{X}$ is used as the model input.

\subsection{Coupled Edge Feature Construction}\label{sec:coupled_feat}

As illustrated in the coupling block of Figure~\ref{fig:framework}, using only topology or only the steady state provides limited temporal cues. We therefore learn an explicit \emph{structure--state coupling} representation from $(G_T,\mathbf{x}_T)$ to improve edge-order discrimination.

To this end, we introduce a graph propagation module~\cite{hamilton2017inductive} that couples structural and state signals at the node level. For each node $i$, the input is the concatenation of its structural statistics and steady state,
\[
\mathbf{h}_i^{(0)} = [\,\text{struct}(i),\, x_i\,],
\]
where $\text{struct}(i)$ includes degree, clustering coefficient, and $k$-core value, and $x_i$ is the steady-state value under the given dynamics. We then apply $L$ layers of neighborhood propagation to obtain coupled node embeddings $\mathbf{h}_i=\mathbf{h}_i^{(L)}$:
\begin{equation}
	\begin{aligned}
		\mathbf{h}_i^{(\ell)}
		&=
		\sigma\!\left(
		W^{(\ell)}_1 \mathbf{h}_i^{(\ell-1)}
		+
		W^{(\ell)}_2
		\sum_{j\in\Gamma(i)}
		\alpha_{ij}\,\mathbf{h}_j^{(\ell-1)}
		\right).
	\end{aligned}
	\label{eq:node_propagation_couple}
\end{equation}
Here, $\ell=1,\dots,L$ denotes the propagation layer, $\alpha_{ij}$ is a topology-based neighborhood normalization coefficient, $W^{(\ell)}_1$ and $W^{(\ell)}_2$ are learnable weight matrices, and $\sigma(\cdot)$ is a nonlinear activation function.

After propagation, $\mathbf{h}_i$ integrates neighborhood information so that the embedding reflects how topology shapes steady-state patterns. We then convert coupled node embeddings into an edge-level coupled feature $\mathbf{f}^{\mathrm{cpl}}(e)$, which serves as the third component of the final edge representation for CPNN (Eq.~\eqref{eq:final_edge_feature}).

Given coupled node embeddings, we construct an edge-level coupled representation for any edge $e=(i,j)$:
\begin{equation}
	\mathbf{f}^{\mathrm{cpl}}(e)
	=
	\big[
	\mathbf{h}_i,\;
	\mathbf{h}_j,\;
	\mathbf{h}_i+\mathbf{h}_j,\;
	|\mathbf{h}_i-\mathbf{h}_j|
	\big],
	\label{eq:edge_representation_couple}
\end{equation}
This formulation captures endpoint commonality, disparity, and their interactions in the coupled space. It implicitly integrates nonlinear structure--state interactions and serves as the input foundation for the subsequent pairwise edge-order predictor.

\subsection{CPNN: Coupled Pairwise Neural Network}

As shown in the CPNN block of Figure~\ref{fig:framework}, the pairwise predictor operates on a unified edge representation that integrates structure, steady state, and their coupling. For each edge $e$, we concatenate $\mathbf{f}^{\mathrm{struct}}(e)$, $\mathbf{f}^{\mathrm{state}}(e)$, and $\mathbf{f}^{\mathrm{cpl}}(e)$ to form
\begin{equation}
	\mathbf{f}(e)
	=
	\big[
	\mathbf{f}^{\mathrm{struct}}(e),\;
	\mathbf{f}^{\mathrm{state}}(e),\;
	\mathbf{f}^{\mathrm{cpl}}(e)
	\big].
	\label{eq:final_edge_feature}
\end{equation}
This representation provides the feature basis for pairwise order prediction.

\subsubsection{Independent Edge Scoring}

CPNN maps each edge to an independent continuous score $z_e$ and compares edges by their scores. This design is permutation-invariant with respect to the input order of an edge pair and naturally supports global ranking at inference time.

For an edge $e$ with input $\mathbf{f}(e)\in\mathbb{R}^{d}$, CPNN uses a multilayer perceptron to output a scalar:
\begin{equation}
	z_e
	=
	\phi\!\big(\mathbf{f}(e)\big),
	\label{eq:cpnn_score}
\end{equation}
where $\phi(\cdot)$ is a learnable nonlinear mapping. A typical instantiation is
\begin{equation}
	z_e
	=
	W_2 \,\sigma\!\left( W_1 \mathbf{f}(e) + \mathbf{b}_1 \right)
	+ \mathbf{b}_2 ,
	\label{eq:cpnn_layers}
\end{equation}
with learnable parameters $W_1,W_2,\mathbf{b}_1,\mathbf{b}_2$ and a nonlinear activation $\sigma(\cdot)$.

\subsubsection{Pairwise Probability Modeling}

Given two edges $e_a$ and $e_b$ with scores $z_a$ and $z_b$, their relative precedence probability is computed by a softmax:
\begin{equation}
	\begin{aligned}
		P(e_a \prec e_b)
		&= \frac{\exp(z_a)}{\exp(z_a)+\exp(z_b)}, \\[4pt]
		P(e_b \prec e_a)
		&= 1 - P(e_a \prec e_b).
	\end{aligned}
	\label{eq:pair_prob}
\end{equation}

We emphasize that the fundamental output of CPNN is the per-edge score $z_e$. Pairwise probabilities are used only for supervision, while the scores $\{z_e\}$ are used downstream to reconstruct a global order.

\subsubsection{Training Objective}

With partially observed formation times, we define a binary label for each supervised edge pair $(e_a,e_b)$:
\begin{equation}
	y_{ab} =
	\begin{cases}
		1, & t(e_a) < t(e_b),\\[2pt]
		0, & t(e_a) > t(e_b),
	\end{cases}
	\label{eq:pair_label}
\end{equation}
indicating whether $e_a$ forms earlier than $e_b$ in the ground truth.

Based on \eqref{eq:pair_label} and \eqref{eq:pair_prob}, we use the pairwise cross-entropy loss:
\begin{equation}
	\begin{aligned}
		\mathcal{L}_{\mathrm{pair}}
		=
		-\!\!\sum_{(a,b)\in\mathcal{P}}
		\Big[
		&\,y_{ab}\log P(e_a \prec e_b) \\
		&+ (1-y_{ab})\log P(e_b \prec e_a)
		\Big].
	\end{aligned}
	\label{eq:pair_loss_formal}
\end{equation}
where $\mathcal{P}$ is the set of supervised pairs.

To improve generalization, we add an $\ell_2$ regularization term:
\begin{equation}
	\mathcal{L}
	=
	\mathcal{L}_{\mathrm{pair}}
	+
	\eta\!\left(
	\Vert W_1\Vert_2^2
	+
	\Vert W_2\Vert_2^2
	+
	\Vert \mathbf{b}_1\Vert_2^2
	+
	\Vert \mathbf{b}_2\Vert_2^2
	\right).
	\label{eq:loss_total_formal}
\end{equation}
where $\eta$ is the regularization coefficient.

After training, each edge is assigned a scalar score $z_e$. We then recover the global edge formation order by aggregating the implied pairwise preferences via Borda count.

\subsection{Global Order Reconstruction}

As shown in the reconstruction block of Figure~\ref{fig:framework}, we convert pairwise precedence predictions into a single global edge sequence. Given the CPNN outputs, we first form pairwise comparisons and then apply a rank-aggregation rule to obtain the final ordering.

\begin{algorithm}[t]
	\renewcommand{\algorithmicrequire}{\textbf{Input:}}
	\renewcommand{\algorithmicensure}{\textbf{Output:}}
		\caption{Borda Aggregation for Global Edge Ordering}
	\label{alg:borda_reconstruction}
	\begin{algorithmic}[1]
		\REQUIRE Pairwise probability matrix 
		\(
		P \in \mathbb{R}^{M \times M},
		\)
		where 
		\(
		P[i][j] = P(e_i \prec e_j)
		\)
		is the predicted probability that edge \(e_i\) is generated earlier than edge \(e_j\).
		\ENSURE Reconstructed generation order \(\hat{T}\).
		
		\STATE Initialize Borda score vector
		\(
		B \in \mathbb{R}^{M}
		\)
		with all zeros.
		
		\FOR{each edge index \(i = 1,\dots,M\)}
		\FOR{each \(j = 1,\dots,M\), \(j \neq i\)}
		\STATE 
		\(
		B[i] \gets B[i] + P[i][j]
		\)
		\ENDFOR
		\ENDFOR
		
		\STATE Obtain global order 
		\(
		\hat{T}
		\)
		by sorting edges in descending order of \(B\):
		\[
		\hat{T} = \operatorname*{argsort}_{i} \big(-B[i]\big).
		\]
		
		\RETURN \(\hat{T}\).
	\end{algorithmic}
\end{algorithm}

After CPNN produces precedence predictions, we aggregate these local preferences into a globally consistent order over edges. Let the final network contain \(M\) edges and denote the pairwise probability matrix by \(P \in \mathbb{R}^{M \times M}\), where \(P[i][j]=P(e_i \prec e_j)\). Borda aggregation assigns each edge a global score by accumulating its expected ``wins'' across all pairwise comparisons and then recovers the overall order by sorting these scores.

Specifically, for each edge \(e_i\), we define its Borda score as
\begin{equation}
	B[i] = \sum_{j \neq i} P[i][j],
	\label{eq:borda_score_simple}
\end{equation}
where a larger score indicates stronger overall precedence in pairwise preferences. Sorting \(B\) in descending order yields the reconstructed sequence \(\hat{T}\). The full procedure is summarized in Algorithm~\ref{alg:borda_reconstruction}.

% !TeX root = TKDE-main.tex
\section{Experiments} \label{sec:exp}

\subsection{Datasets}

Our experiments use (i) real temporal networks with ground-truth edge formation times and (ii) corresponding steady-state observations generated by simulating network dynamics on the final snapshot. We first describe the temporal networks and their statistics, and then introduce the dynamical processes used to obtain steady states.

\subsubsection{Temporal Coauthorship Networks}

We construct temporal coauthorship networks from the American Physical Society (APS) publication records, following prior work~\cite{wang2024reconstructing,hu2014conditions}. We consider six PACS subfields: ComplexNet, Chaos, Fluctuations, Interfaces, PhaseTrans, and Thermodynamics.

\begin{table}[t]
		\centering
		\caption{Dataset statistics.}
		\label{tab:dataset_stats}
		\scalebox{1}{
		\begin{tabular}{lccccc}
		\toprule
		\textbf{Network} & \(\mathbf{N}\) & \(\mathbf{E}\) & \(\mathbf{E_d}\) & \(\mathbf{P_{E_d}}\) & \(\mathbf{S}\) \\
		\midrule
			ComplexNet       & 225   & 413   & 84,445       & 0.9926 & 172   \\
			Chaos            & 2,055 & 3,758 & 7,048,311    & 0.9984 & 1,118 \\
			Fluctuations     & 1,248 & 2,198 & 2,408,204    & 0.9974 & 731   \\
			Interfaces       & 2,745 & 6,718 & 22,506,659   & 0.9975 & 1,040 \\
			PhaseTrans       & 1,113 & 1,882 & 1,764,785    & 0.9970 & 654   \\
			Thermodynamics   & 158   & 228   & 25,668       & 0.9919 & 131   \\
			\bottomrule
		\end{tabular}
		}
	\end{table}

In each subfield, nodes represent authors, and an undirected edge records the first year that two authors coauthor a paper. Sorting edges by their first-occurrence years yields the ground-truth formation order. We use the final snapshot \(G_T\) for feature extraction and steady-state simulation.
Table~\ref{tab:dataset_stats} reports dataset statistics, including the numbers of nodes \(N\) and edges \(E\), the number of yearly snapshots \(S\), and the number of distinguishable edge pairs \(E_d\). Here, \(E_d\) counts edge pairs with different ground-truth times (thus providing valid supervision), and \(P_{E_d}=E_d/\binom{E}{2}\) is the fraction of supervised pairs among all possible pairs.

\subsubsection{Dynamical Processes and Steady-State Generation}

To obtain steady-state observations aligned with the final snapshot $G_T$, we simulate three representative dynamics on $G_T$: SIS~\cite{murphy2021deep}, Gene~\cite{badia2023gene}, and Opinion~\cite{cao2024discrete}. Each process starts from $\mathbf{x}(0)\sim\mathrm{Unif}(0,1)^N$ and is iterated until convergence ($\|\mathbf{x}(t{+}1)-\mathbf{x}(t)\|_2 < 10^{-6}$) or 1000 steps. The converged state $\mathbf{x}_T$ is used as the steady state.

\textit{SIS epidemic dynamics.}
We use the susceptible--infected--susceptible (SIS) model with infection and recovery rates $\beta=0.4$ and $\delta=0.3$. Let $p_i(t)\in[0,1]$ be the infection probability. The discrete-time update is
\begin{equation}
	\begin{aligned}
		p_i(t{+}1)
		&= (1-\delta)\,p_i(t)  \\
		&\quad {}+ (1-p_i(t))
		\Bigl[1 - \!\!\prod_{j\in\Gamma(i)} (1-\beta A_{ij} p_j(t))\Bigr].
	\end{aligned}
\end{equation}
The process converges to an endemic steady state.

\medskip
\textit{Gene regulation dynamics.}
We simulate gene expression under mutual regulation using a Hill-type update. Each node has parameters $(b_{1,i}, b_{2,i})$ sampled from $[0,1]$ and $[0.5,1.5]$. Let
\begin{equation}
	I_i(t) = \sum_{j\in\Gamma(i)} A_{ij}\,x_j(t)
\end{equation}
be the total regulatory input, and update
\begin{equation}
	x_i(t+1)
	=
	b_{1,i}
	+
	b_{2,i}\,
	\frac{I_i(t)^n}{1 + I_i(t)^n},
\end{equation}
where $n>1$ is the Hill coefficient.

\medskip
\textit{Opinion dynamics.}
We use a heterogeneous DeGroot-type update to model interpersonal influence on continuous opinions. Each node $i$ has a susceptibility parameter $\theta_i$ sampled from $[1,1.5]\cup[3.5,4]$. Let $x_i(t)$ be the opinion of node $i$ at time $t$ and define the neighborhood mean
\begin{equation}
	m_i(t) = \frac{1}{k(i)} \sum_{j\in\Gamma(i)} A_{ij}\,x_j(t),
\end{equation}
then update
\begin{equation}
	x_i(t+1) = x_i(t) + \theta_i\bigl(m_i(t) - x_i(t)\bigr).
\end{equation}
The process converges to a steady state $\mathbf{x}_T$.

%\subsubsection{使用模拟动力学稳态是否合理？（需要专门小段说明）}
%
%关键论点：
%\begin{itemize}
%	\item 目标不是恢复真实动力学，而是验证“稳态 $\rightarrow$ 结构演化”的信息性
%	\item 不同动力学生成的稳态都保留结构演化信号，说明方法对动力学选择具有稳健性
%	\item 文献支持：如 NC 刘博等高质量工作均采用模拟动力学进行验证
%	\item 高质量工作普遍是“预测目标为结构/演化，而非动力学本身”
%\end{itemize}

\begin{figure*}[htbp]
	\centering
\includegraphics[width=0.85\textwidth]{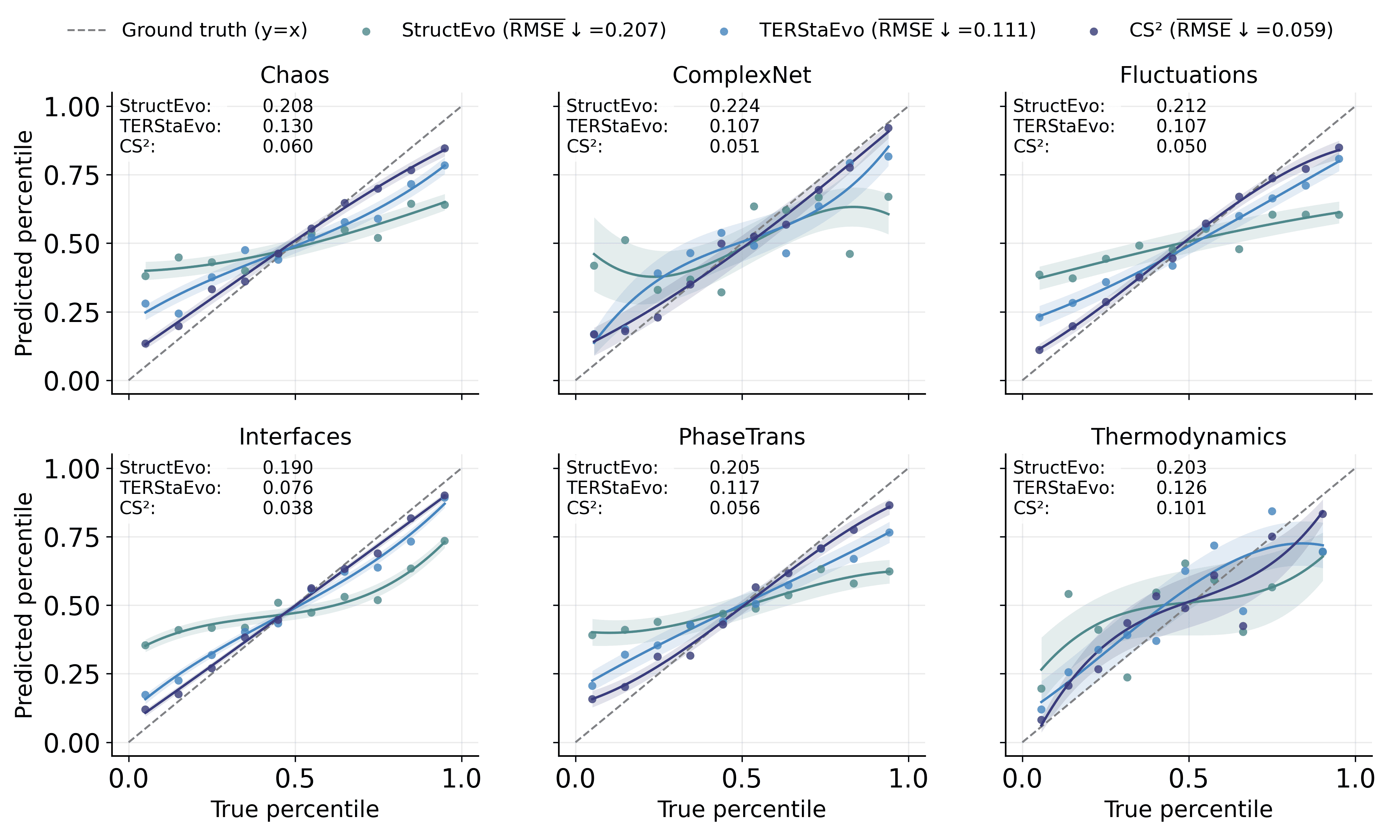}
	\vspace{-0.25cm}
	\caption{Global ordering consistency under SIS dynamics.
		Each panel shows the relationship between the true normalized edge rank and the predicted normalized edge rank on one dataset.}
	\label{fig:rank_scatter_example}
\end{figure*}

\subsection{Baselines}

We compare CS$^2$ with representative baselines that cover topology-driven evolution reconstruction and structural embedding methods. To isolate the effect of steady-state information, we also include state-augmented variants that concatenate steady-state features with the corresponding structural edge representations.

\begin{itemize}
	\item \textit{StructEvo.}
	A representative topology-driven framework (Nature Communications) that infers edge formation order purely from the final snapshot by learning time-discriminative structural patterns under partial timestamp supervision~\cite{wang2024reconstructing}.

	\item \textit{StructStaEvo.}
	A state-augmented variant of \textit{StructEvo} that concatenates handcrafted structural features with steady-state features at the edge level~\cite{wang2024reconstructing}.

	\item \textit{TopoDiff.}
	A structure-driven method designed for cross-network transfer. It jointly trains on multiple temporal networks and uses diffusion-style temporal augmentation to improve generalization to unseen static networks~\cite{Xu2026TKDE}.

	\item \textit{TopoDiffStaEvo.}
	A state-augmented variant of \textit{TopoDiff} that concatenates steady-state features with its topology-based edge representation~\cite{Xu2026TKDE}.

	\item \textit{N2VecEvo.}
	Node2Vec learns node embeddings on the final snapshot; we then construct edge representations from endpoint embeddings for order inference~\cite{grover2016node2vec}.

	\item \textit{N2VecStaEvo.}
	A state-augmented variant of \textit{N2VecEvo} by concatenating Node2Vec-based edge representations with steady-state features~\cite{grover2016node2vec}.

	\item \textit{LLaCEEvo.}
	A structural embedding baseline that learns representations on the final snapshot and constructs edge features for order inference~\cite{liu2024llace}.

	\item \textit{LLaCEStaEvo.}
	A state-augmented variant of \textit{LLaCEEvo} by concatenating structural embeddings with steady-state features~\cite{liu2024llace}.

	\item \textit{TEREvo.}
	A structural embedding baseline that learns representations on the final snapshot and performs edge-order inference from the resulting edge features~\cite{wang2023efficient}.

	\item \textit{TERStaEvo.}
	A state-augmented variant of \textit{TEREvo} by concatenating TER-based edge representations with steady-state features~\cite{wang2023efficient}.
\end{itemize}

\subsection{Overall Pairwise Ranking Accuracy}

\begin{table*}[t]
	\centering
	\caption{Overall pairwise ranking accuracy on six real-world datasets under different dynamics.
		Higher is better. The best and second-best results under each setting are highlighted in
		\textbf{bold} and \underline{underline}, respectively.}
	\label{tab:pairwise_accuracy}
	\renewcommand{\arraystretch}{1.2}
	\setlength{\tabcolsep}{6pt}
	
		\begin{tabular}{llcccccc>{\columncolor[gray]{0.9}}c}
			\toprule
			\textbf{Dynamics} & \textbf{Method} & \textbf{Chaos} & \textbf{Complex} & \textbf{Fluctuations} & \textbf{Interfaces} & \textbf{Phase} & \textbf{Thermo} & \textbf{Avg} \\
			\midrule
				\multirow{11}{*}{\textit{SIS}}
				& StructEvo     & 0.5964 & 0.5970 & 0.5949 & 0.6199 & 0.5999 & 0.6449 & 0.6088 \\
				& StructStaEvo  & 0.6105 & 0.5288 & 0.6138 & 0.6359 & 0.5866 & 0.6176 & 0.5989 \\
				& TopoDiff      & 0.5677 & 0.5353 & 0.5597 & 0.5583 & 0.5712 & 0.6576 & 0.5750 \\
				& TopoDiffStaEvo & 0.5709 & 0.5979 & 0.5566 & 0.5702 & 0.5579 & 0.6693 & 0.5871 \\
				& N2VecEvo      & 0.7740 & \underline{0.8131} & 0.8003 & 0.8220 & 0.7352 & 0.7649 & 0.7849 \\
			& N2VecStaEvo   & \underline{0.7814} & 0.7864 & \underline{0.8115} & \underline{0.8348} & 0.7619 & 0.7327 & 0.7848 \\
			& LLaCEEvo      & 0.7521 & 0.8036 & 0.7743 & 0.8266 & 0.7706 & 0.7551 & 0.7804 \\
			& LLaCEStaEvo   & 0.7447 & 0.7866 & 0.8020 & 0.8339 & \underline{0.7754} & \underline{0.7746} & 0.7862 \\
			& TEREvo        & 0.7335 & 0.8047 & 0.7494 & 0.7982 & 0.7410 & 0.6702 & 0.7495 \\
			& TERStaEvo     & 0.7061 & 0.7662 & 0.7385 & 0.7984 & 0.7214 & 0.7502 & 0.7468 \\
		& \textbf{CS$^2$}
		& \makecell{\textbf{0.8135} \\ (+4.1\%)}
		& \makecell{\textbf{0.8558} \\ (+5.3\%)}
		& \makecell{\textbf{0.8457} \\ (+4.2\%)}
		& \makecell{\textbf{0.8664} \\ (+3.8\%)}
			& \makecell{\textbf{0.8266} \\ (+6.6\%)}
			& \makecell{\textbf{0.8185} \\ (+5.7\%)}
			& \textbf{0.8378} \\
				\midrule
				\multirow{11}{*}{\textit{Gene}}    
				& StructEvo     & 0.5964 & 0.5970 & 0.5949 & 0.6199 & 0.5999 & 0.6449 & 0.6088 \\
				& StructStaEvo  & 0.5917 & 0.5531 & 0.5870 & 0.6235 & 0.5979 & 0.5854 & 0.5898 \\
				& TopoDiff      & 0.5677 & 0.5353 & 0.5597 & 0.5583 & 0.5712 & 0.6576 & 0.5750 \\
				& TopoDiffStaEvo & 0.5528 & 0.5682 & 0.5607 & 0.5737 & 0.5645 & 0.6312 & 0.5752 \\
				& N2VecEvo      & 0.7740 & \underline{0.8131} & 0.8003 & 0.8220 & 0.7352 & 0.7649 & 0.7849 \\
				& N2VecStaEvo   & \underline{0.7814} & 0.7864 & \underline{0.8115} & \underline{0.8348} & 0.7619 & 0.7327 & 0.7848 \\
				& LLaCEEvo      & 0.7521 & 0.8036 & 0.7743 & 0.8266 & 0.7706 & 0.7551 & 0.7804 \\
				& LLaCEStaEvo   & 0.7447 & 0.7866 & 0.8020 & 0.8339 & \underline{0.7754} & \underline{0.7746} & 0.7862 \\
				& TEREvo        & 0.7335 & 0.8047 & 0.7494 & 0.7982 & 0.7410 & 0.6702 & 0.7495 \\
			& TERStaEvo     & 0.7061 & 0.7662 & 0.7385 & 0.7984 & 0.7214 & 0.7502 & 0.7468 \\
			& \textbf{CS$^2$}
			& \makecell{\textbf{0.8017} \\ (+2.6\%)}
			& \makecell{\textbf{0.8380} \\ (+3.1\%)}
			& \makecell{\textbf{0.8361} \\ (+3.0\%)}
			& \makecell{\textbf{0.8723} \\ (+4.5\%)}
			& \makecell{\textbf{0.8160} \\ (+5.2\%)}
			& \makecell{\textbf{0.8078} \\ (+4.3\%)}
			& \textbf{0.8287} \\
				\midrule
				\multirow{11}{*}{\textit{Opinion}}
				& StructEvo     & 0.5964 & 0.5970 & 0.5949 & 0.6199 & 0.5999 & 0.6449 & 0.6088 \\
				& StructStaEvo  & 0.6080 & 0.5484 & 0.6408 & 0.6310 & 0.6591 & 0.6595 & 0.6245 \\
				& TopoDiff      & 0.5677 & 0.5353 & 0.5597 & 0.5583 & 0.5712 & 0.6576 & 0.5750 \\
				& TopoDiffStaEvo & 0.5605 & 0.6024 & 0.5427 & 0.5609 & 0.5739 & 0.6585 & 0.5832 \\
				& N2VecEvo      & 0.7740 & \underline{0.8131} & 0.8003 & 0.8220 & 0.7352 & 0.7649 & 0.7849 \\
				& N2VecStaEvo   & \underline{0.7814} & 0.7864 & \underline{0.8115} & \underline{0.8348} & 0.7619 & 0.7327 & 0.7848 \\
				& LLaCEEvo      & 0.7521 & 0.8036 & 0.7743 & 0.8266 & 0.7706 & 0.7551 & 0.7804 \\
				& LLaCEStaEvo   & 0.7447 & 0.7866 & 0.8020 & 0.8339 & \underline{0.7754} & \underline{0.7746} & 0.7862 \\
				& TEREvo        & 0.7335 & 0.8047 & 0.7494 & 0.7982 & 0.7410 & 0.6702 & 0.7495 \\
			& TERStaEvo     & 0.7061 & 0.7662 & 0.7385 & 0.7984 & 0.7214 & 0.7502 & 0.7468 \\
			& \textbf{CS$^2$}
			& \makecell{\textbf{0.7995} \\ (+2.3\%)}
			& \makecell{\textbf{0.8401} \\ (+3.3\%)}
			& \makecell{\textbf{0.8252} \\ (+1.7\%)}
			& \makecell{\textbf{0.8650} \\ (+3.6\%)}
			& \makecell{\textbf{0.8027} \\ (+3.5\%)}
			& \makecell{\textbf{0.8137} \\ (+5.0\%)}
			& \textbf{0.8244} \\
			\bottomrule
		\end{tabular}
	\end{table*}

Table~\ref{tab:pairwise_accuracy} reports the overall pairwise ranking accuracy, i.e., the probability of correctly predicting the temporal precedence for a randomly sampled edge pair. This metric directly evaluates the core capability required by our framework: producing reliable local pairwise comparisons that can be aggregated into a global evolution order.

Overall, \textit{CS$^2$} achieves the best performance across all datasets and all dynamical settings, with average accuracies of \textit{0.8378} (SIS), \textit{0.8287} (Gene), and \textit{0.8244} (Opinion). Across datasets, the strongest competitors are typically structural-embedding baselines (\textit{N2VecEvo}, \textit{LLaCEEvo}, and \textit{TEREvo}) and their state-augmented variants. Importantly, \textit{CS$^2$} improves over the strongest baseline in \emph{every} dataset--dynamics setting, by 4.0\% on average and up to 6.6\% (parentheses in Table~\ref{tab:pairwise_accuracy}), indicating that explicitly learning structure--state coupling provides consistent benefits beyond naive feature augmentation.

\subsection{Global Evolution Consistency}

We next evaluate whether local pairwise predictions can be aggregated into a \emph{globally consistent} evolution sequence. While pairwise accuracy measures correctness on edge pairs, it does not guarantee that the induced full order is coherent. We therefore reconstruct a complete edge sequence via Borda aggregation and quantify its agreement with the ground truth using Spearman's rank correlation coefficient $\rho$.
As shown in Table~\ref{tab:spearman_rho_sis}, under \textit{SIS} dynamics, \textit{CS$^2$} achieves the best global consistency on all datasets (Avg $\rho=\textit{0.8345}$), improving over the strongest baseline by 4.7\%--15.1\%. Results under \textit{Gene} and \textit{Opinion} dynamics are reported in Appendix Table~\ref{tab:spearman_rho_gene_opinion}.

\begin{table*}[t]
	\centering
	\caption{Global ordering consistency (Spearman-$\rho$) under \textit{SIS} dynamics. Higher is better.}
	\label{tab:spearman_rho_sis}
	\renewcommand{\arraystretch}{1.2}
	\setlength{\tabcolsep}{6pt}
	
		\begin{tabular}{lcccccc>{\columncolor[gray]{0.9}}c}
			\toprule
			\textbf{Method} & \textbf{Chaos} & \textbf{Complex} & \textbf{Fluctuations} & \textbf{Interfaces} & \textbf{Phase} & \textbf{Thermo} & \textbf{Avg} \\
			\midrule
			StructEvo     & 0.2831 & 0.2749 & 0.2786 & 0.3489 & 0.2909 & 0.4016 & 0.3130 \\
			StructStaEvo  & 0.3236 & 0.0647 & 0.3308 & 0.3940 & 0.2530 & 0.3264 & 0.2821 \\
			TopoDiff      & 0.2012 & 0.1216 & 0.1764 & 0.1723 & 0.2149 & 0.4582 & 0.2241 \\
			TopoDiffStaEvo & 0.2103 & 0.2799 & 0.1670 & 0.2074 & 0.1724 & 0.5195 & 0.2594 \\
			N2VecEvo      & 0.7309 & \underline{0.7885} & 0.7727 & 0.8234 & 0.6176 & 0.7008 & 0.7390 \\
			N2VecStaEvo   & \underline{0.7367} & 0.7411 & \underline{0.7955} & \underline{0.8420} & 0.6802 & 0.6234 & 0.7365 \\
			LLaCEEvo      & 0.6687 & 0.7681 & 0.7136 & 0.8182 & 0.7071 & 0.6757 & 0.7252 \\
			LLaCEStaEvo   & 0.6460 & 0.7467 & 0.7694 & 0.8282 & \underline{0.7098} & \underline{0.7084} & 0.7348 \\
			TEREvo        & 0.6183 & 0.7699 & 0.6584 & 0.7394 & 0.6266 & 0.4612 & 0.6456 \\
			TERStaEvo     & 0.5574 & 0.6871 & 0.6356 & 0.7541 & 0.6041 & 0.6512 & 0.6482 \\
		\textbf{CS$^2$}
		& \makecell{\textbf{0.7977} \\ (+8.3\%)}
		& \makecell{\textbf{0.8626} \\ (+9.4\%)}
		& \makecell{\textbf{0.8450} \\ (+6.2\%)}
		& \makecell{\textbf{0.8813} \\ (+4.7\%)}
		& \makecell{\textbf{0.8170} \\ (+15.1\%)}
		& \makecell{\textbf{0.8036} \\ (+13.4\%)}
		& \textbf{0.8345} \\
		\bottomrule
	\end{tabular}
\end{table*}
To visualize global consistency across evolution stages, we partition edges into 10 percentile bins by their ground-truth ranks and plot, for each bin, the median predicted rank with an intra-bin variance band. Curves closer to the ideal $y=x$ line indicate better global agreement, while narrower bands indicate more stable predictions. Figure~\ref{fig:rank_scatter_example} shows the binned plots under \textit{SIS} dynamics; the corresponding plots for \textit{Gene} and \textit{Opinion} are provided in Appendix~\ref{apdx:rank_scatter_gene_opinion}.

All methods exhibit an increasing trend, but \textit{CS$^2$} stays closer to the $y=x$ line with tighter variance bands, indicating more accurate and stable global recovery across evolution stages, consistent with the Spearman-$\rho$ results in Table~\ref{tab:spearman_rho_sis}.

\begin{figure*}[!t]
	\centering
\includegraphics[width=0.93\textwidth]{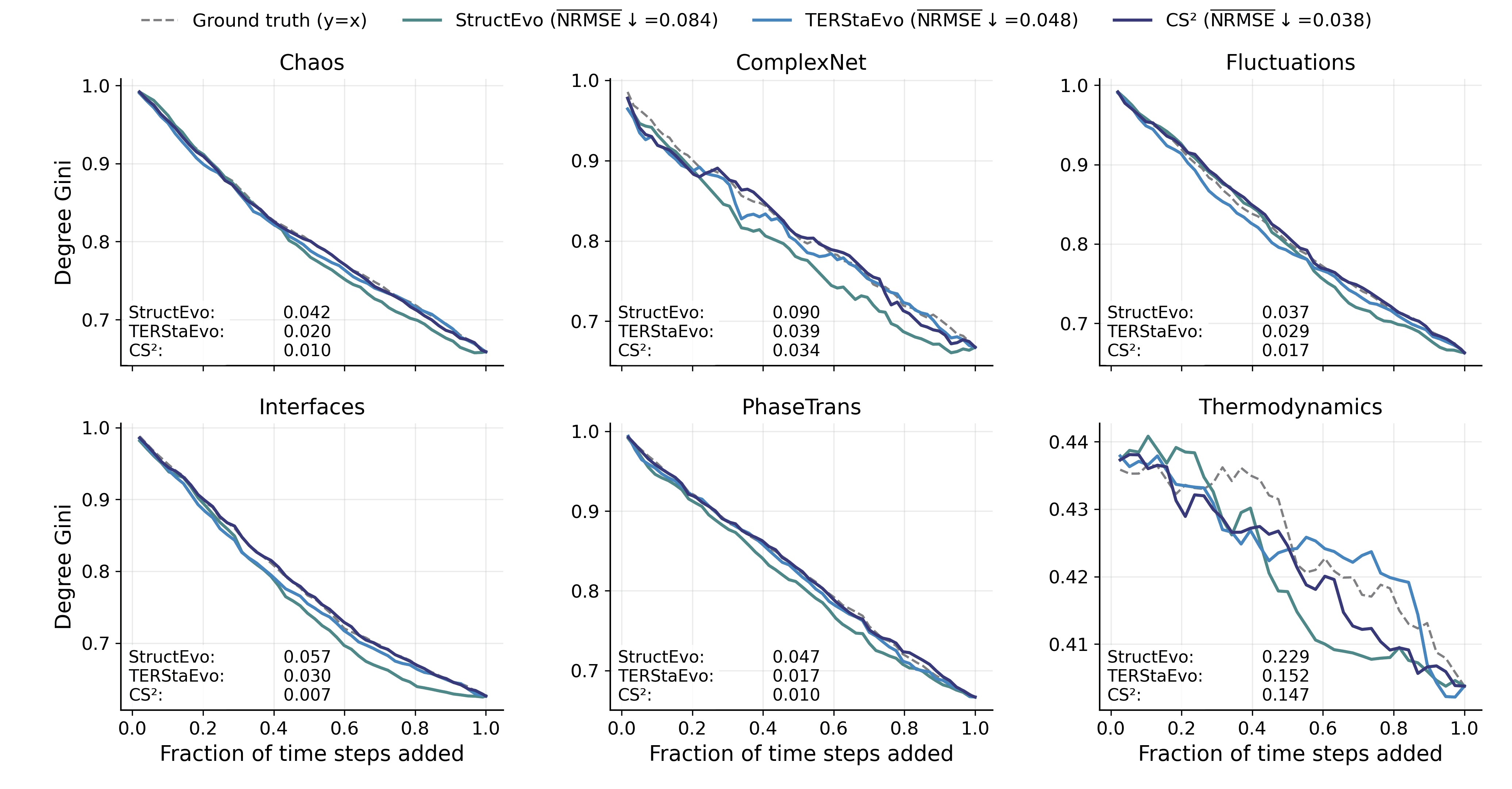}
	\vspace{-0.25cm}
	\caption{Evolution of degree Gini across datasets under SIS dynamics.}
	\vspace{-0.18cm}
	\label{fig:degree_gini_evo}
\end{figure*}

\begin{figure*}[!t]
	\centering
\includegraphics[width=0.93\textwidth]{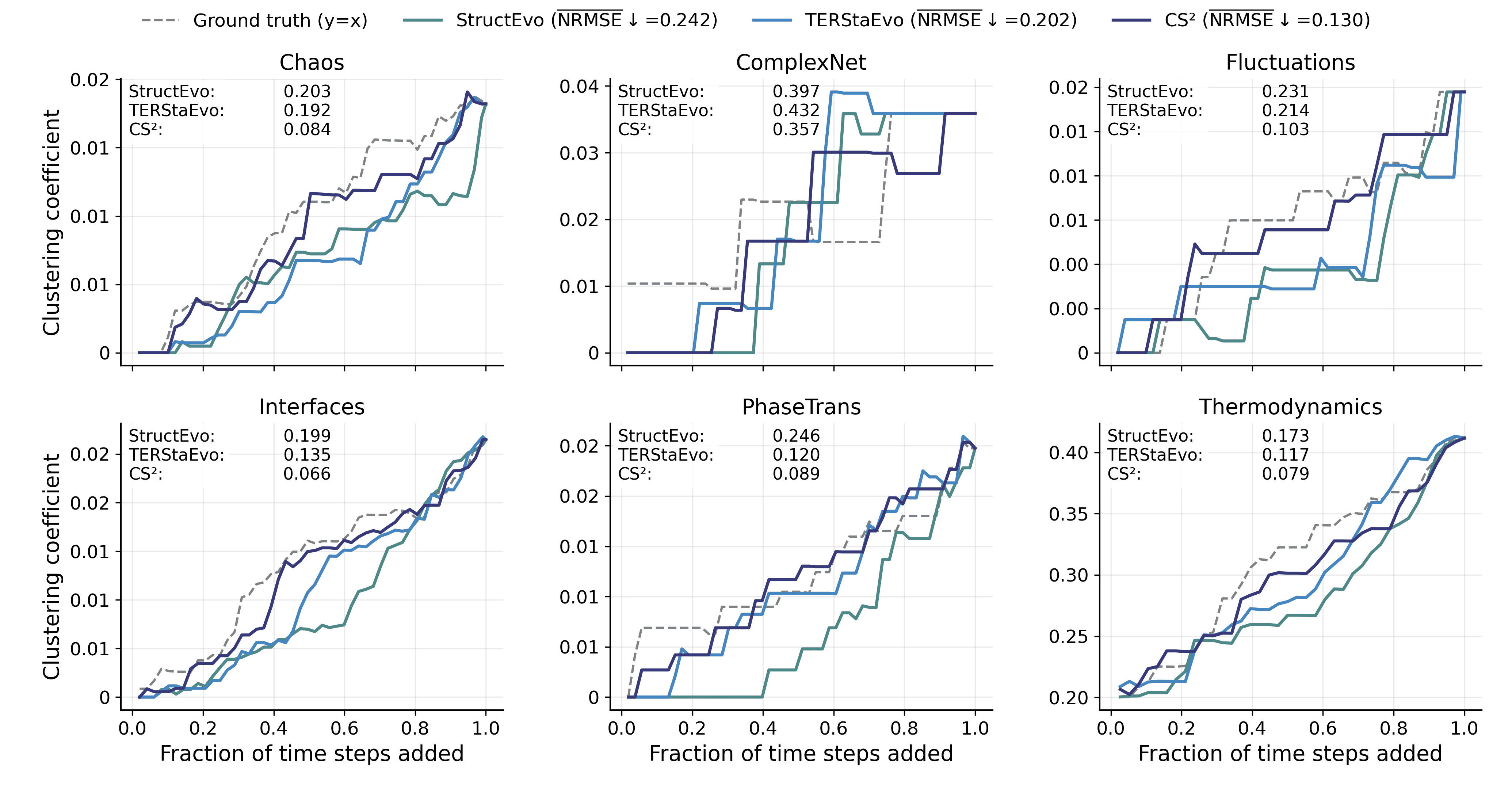}
	\vspace{-0.25cm}
	\caption{Evolution of average clustering coefficient across datasets under SIS dynamics.}
	\vspace{-0.2cm}
	\label{fig:clustering_evo}
\end{figure*}

\begin{figure*}[!t]
	\centering
\includegraphics[width=0.93\textwidth]{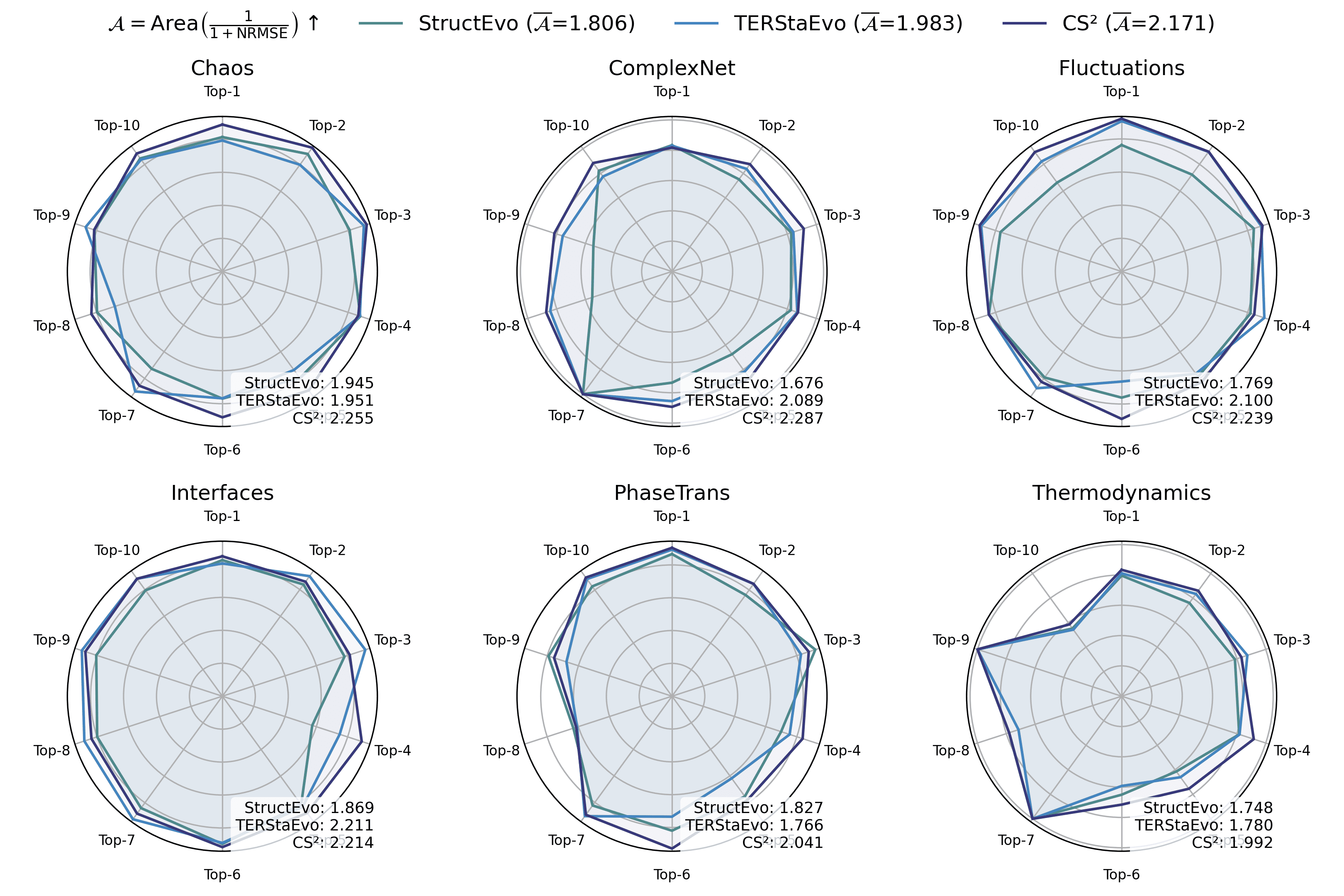}
	\vspace{-0.25cm}
	\caption{Top-hub degree evolution fidelity across datasets under SIS dynamics.}
	\vspace{-0.18cm}
	\label{fig:hubs_radar}
\end{figure*}

\subsection{Global Property Recovery}

Pairwise and global ranking metrics assess ordering correctness, but they do not directly reveal whether the inferred history can recover the network's \emph{evolutionary process} in a functionally meaningful way. To bridge this gap, we evaluate global property recovery: given a predicted ordering, we add edges sequentially to obtain intermediate graphs $\{G_t\}$ and compare the resulting macroscopic trajectories with those induced by the ground-truth order. This directly tests whether the recovered history reproduces how characteristic structures (e.g., clustering and hubs) emerge and evolve over time.
We consider three representative measures: (i) \textit{Clustering}, the average clustering coefficient of $G_t$ (triangle closure and clustered-structure formation); (ii) \textit{Degree Gini}, the Gini coefficient of degrees (the emergence of degree heterogeneity and hub concentration); and (iii) \textit{Hub radar}, which summarizes Top-$K$ hub trajectory fidelity by converting hub-level NRMSE to $s=1/(1+\mathrm{NRMSE})$ and aggregating by the radar-area score $\mathcal{A}$. For \textit{Clustering} and \textit{Degree Gini}, we quantify trajectory-level deviation by the NRMSE between the predicted and ground-truth curves.

Figures~\ref{fig:degree_gini_evo}--\ref{fig:hubs_radar} report results under \textit{SIS} dynamics; results for \textit{Gene} and \textit{Opinion} are provided in Appendix~\ref{apdx:global_property_gene_opinion}. Overall, \textit{CS$^2$} most faithfully reproduces the ground-truth evolution trajectories across all three measures, yielding curves closer to the ground truth throughout the evolution. For \textit{Degree Gini}, \textit{CS$^2$} achieves the lowest average error (NRMSE$\downarrow$: 0.038 vs.\ 0.048 for \textit{TERStaEvo} and 0.084 for \textit{StructEvo}), indicating a more accurate recovery of how degree heterogeneity emerges over time. For \textit{Clustering}, \textit{CS$^2$} also achieves the lowest error (NRMSE$\downarrow$: 0.130 vs.\ 0.202/0.242), suggesting better reconstruction of triangle closure and clustered-structure formation. For \textit{Hub radar}, \textit{CS$^2$} obtains the highest area-based similarity score ($\overline{\mathcal{A}}{=}2.171$ vs.\ 1.983/1.806) and maintains larger radar coverage across datasets and hub ranks, reflecting higher fidelity in the growth trajectories of top hubs. Overall, accurate history inference enables faithful recovery of macroscopic network evolution.

\begin{table*}[t]
	\centering
	\caption{Pairwise ranking accuracy of \textit{StateEvo} across datasets under different dynamics. Higher is better.}
	\label{tab:state_only_performance}
	\renewcommand{\arraystretch}{1.15}
	\setlength{\tabcolsep}{6pt}
	
	\begin{tabular}{lcccccc>{\columncolor[gray]{0.9}}c}
		\toprule
		\textbf{Dynamics} & \textbf{Chaos} & \textbf{Complex} & \textbf{Fluctuations} & \textbf{Interfaces} & \textbf{Phase} & \textbf{Thermo} & \textbf{Avg} \\
		\midrule
		Gene    & 0.5610 & 0.5825 & 0.5490 & 0.5564 & 0.5328 & 0.6683 & 0.5750 \\
		Opinion & 0.5541 & 0.6184 & 0.5386 & 0.5369 & 0.5791 & 0.6293 & 0.5761 \\
		SIS     & 0.5688 & 0.4843 & 0.5350 & 0.5516 & 0.5446 & 0.6312 & 0.5526 \\
		\bottomrule
	\end{tabular}
\end{table*}

\subsection{Steady-State-Only Evolution Inference}\label{sec:state_only}

We evaluate a steady-state-only setting, \textit{StateEvo}, which predicts pairwise temporal precedence using edge features constructed solely from the final steady state $\mathbf{x}_T$ (without any structural inputs). Table~\ref{tab:state_only_performance} reports the corresponding pairwise ranking accuracy under three dynamics. Overall, \textit{StateEvo} attains stable average accuracy (0.55--0.58), indicating that steady states provide a nontrivial and independent signal for temporal ordering when reliable topology is limited or unavailable. Performance varies across datasets (e.g., consistently higher on Thermodynamics and lower on ComplexNet under SIS), suggesting that identifiability depends on both topology and the dynamical mechanism. Taken together, these results highlight the practical value of steady-state observations as an alternative information source under structure-limited settings.

\begin{figure}[t]
	\centering
	\includegraphics[width=0.95\linewidth]{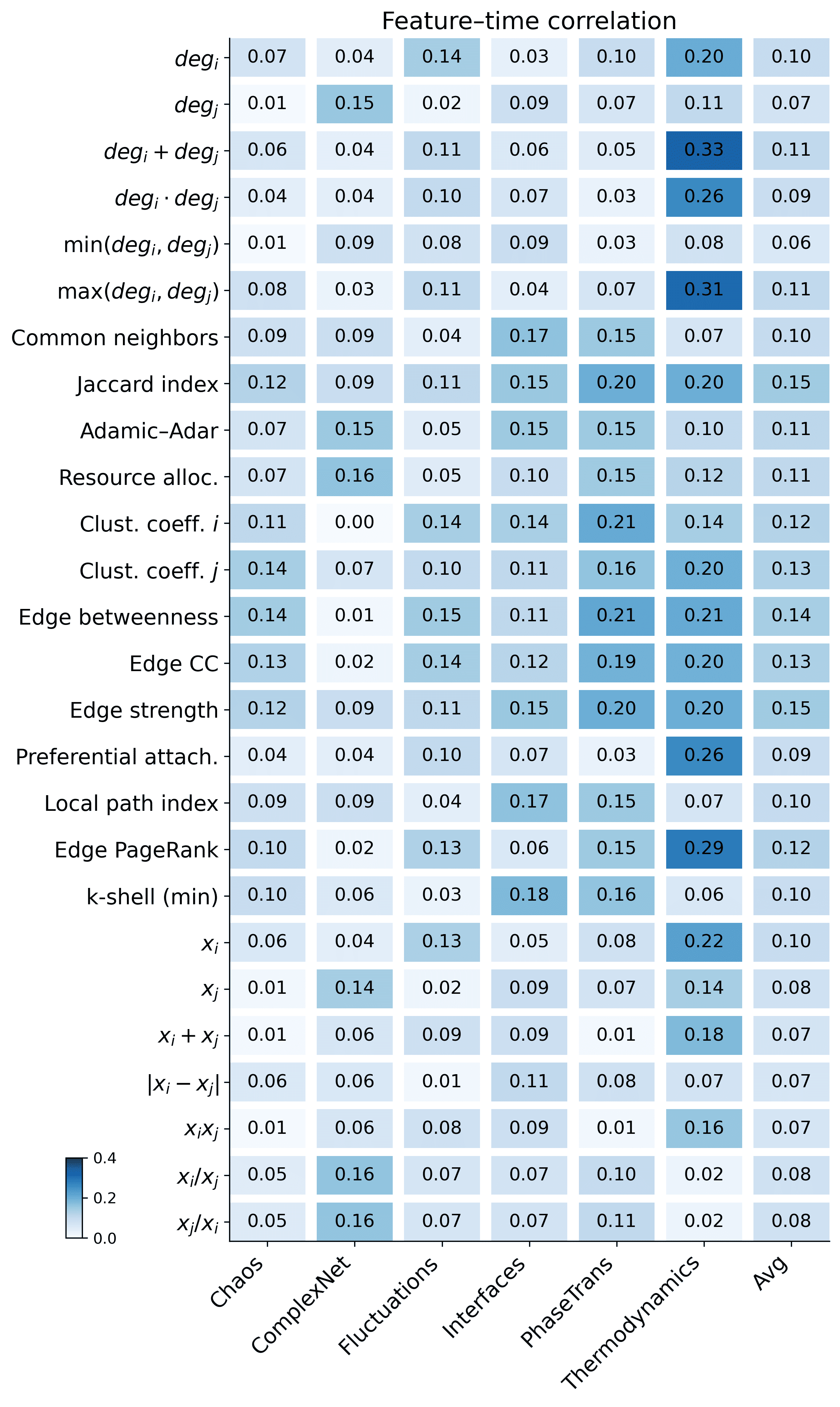}
	\vspace{-0.2cm}
	\caption{Heatmap of feature–time Spearman correlations.}
	\label{fig:feature_importance}
\end{figure}

\subsection{Feature Importance Analysis}

We quantify the association between each individual feature and the ground-truth edge formation time using Spearman correlation. As shown in Figure~\ref{fig:feature_importance}, most single features have weak correlations (typically $<0.2$), and only a few reach around $0.25$ on certain datasets. This indicates that temporal cues are fragmented in the final snapshot and dispersed across multiple structural relations and steady-state patterns, rather than being dominated by any single indicator. It therefore motivates learning coupled representations that integrate complementary signals across structure and steady state.

\section{Conclusion}
This paper studied the problem of recovering edge formation order from a single final network snapshot by leveraging steady-state observations as an additional and often more accessible signal beyond topology. We proposed CS$^2$, a dynamics-informed framework that explicitly learns structure--state coupled edge representations for temporal ordering. Extensive experiments on real temporal networks under multiple dynamical processes demonstrated that CS$^2$ consistently outperforms strong baselines, improving pairwise edge precedence accuracy by 4.0\% on average and global ordering consistency (Spearman-$\rho$) by 7.7\% on average, while more faithfully recovering macroscopic evolution trajectories such as clustering formation, degree heterogeneity, and hub growth. Importantly, the results also indicate that steady-state features alone provide nontrivial discriminative power, highlighting the practicality of evolution inference in settings where reliable network structure is incomplete or unavailable.

\bibliographystyle{IEEEtran}
\bibliography{tkde}

\vspace*{-1.1cm}
\begin{IEEEbiography}[{\includegraphics[width=1in,height=1.25in,clip,keepaspectratio]{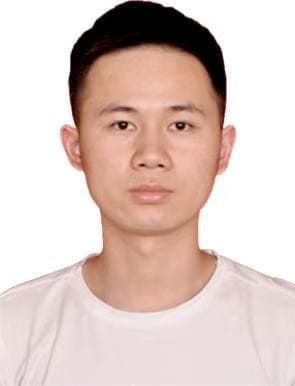}}]{En Xu}
received the BS and PhD degrees in Computer Science and Technology from Northwestern Polytechnical University, Xi'an, China, in 2018 and 2023, respectively. He was a postdoctoral research fellow at Hong Kong Baptist University from 2023 to 2024 and is currently a postdoctoral research fellow with the Department of Electronic Engineering, Tsinghua University, Beijing, China. His research interests include the predictability of time series and complex networks, with a current focus on AI for complex networks.
\end{IEEEbiography}

\vspace*{-1.1cm}
\begin{IEEEbiography}[{\includegraphics[width=1in,height=1.25in,clip,keepaspectratio]{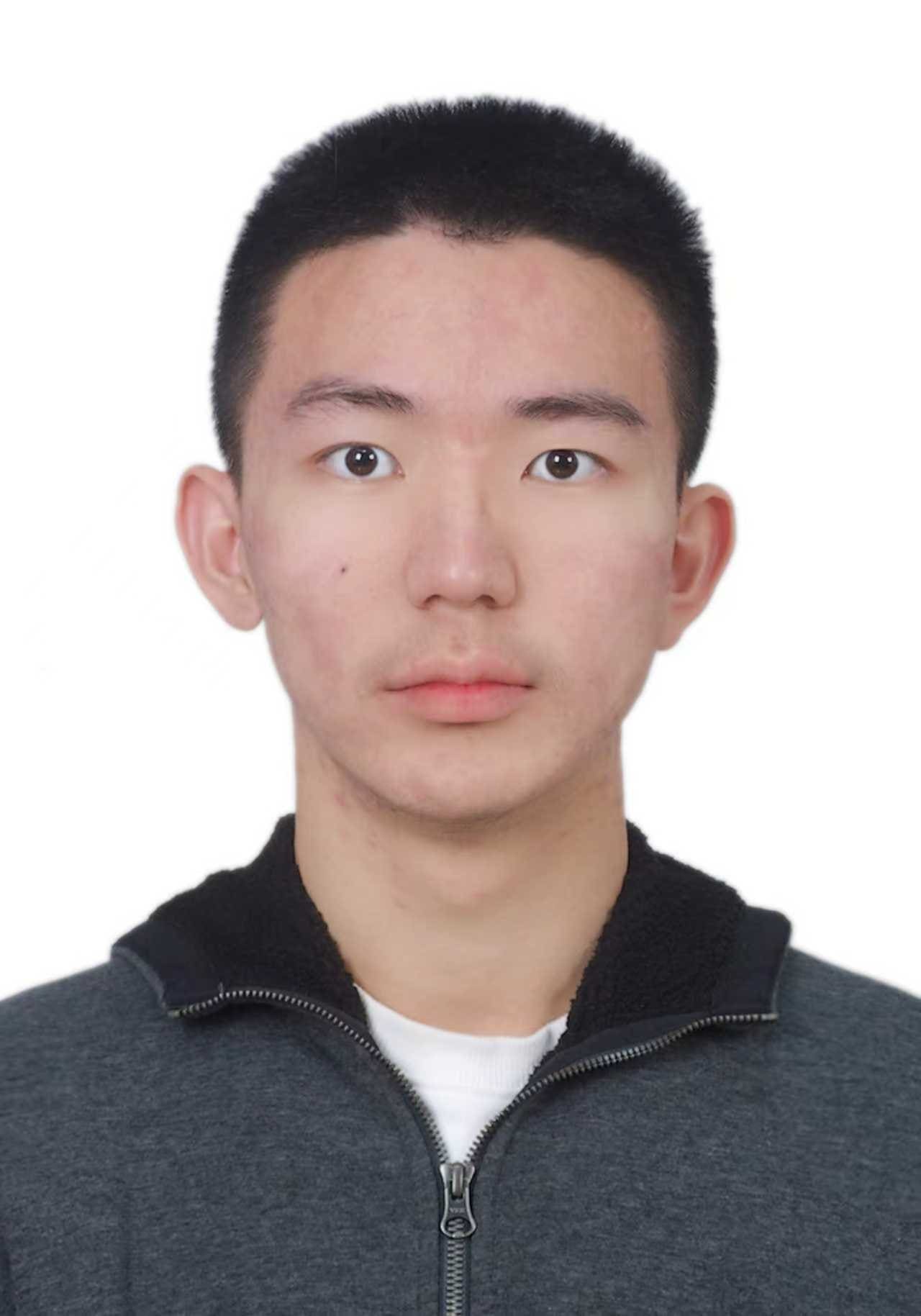}}]{Shihe Zhou}
is currently pursuing the B.S. degree in mechanics at the Tsien Excellence in Engineering Programme (TEEP), Tsinghua University, Beijing, China. His research interests include generative AI for dynamics and graph dynamics prediction.
\end{IEEEbiography}

\vspace*{-1.1cm}
\begin{IEEEbiography}[{\includegraphics[width=1in,height=1.25in,clip,keepaspectratio]{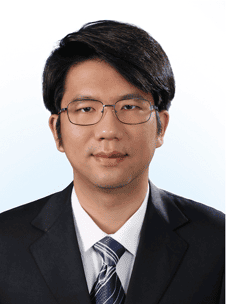}}]{Huandong Wang}
received the BS degree in electronic engineering and the second BS degree in mathematical sciences from Tsinghua University, Beijing, China, in 2014 and 2015, respectively, and the PhD degree in electronic engineering from Tsinghua University, Beijing, China, in 2019. He is currently a research associate with the Department of Electronic Engineering, Tsinghua University. His research interests include urban computing, mobile Big Data mining, and machine learning.
\end{IEEEbiography}

\vspace*{-1.1cm}
\begin{IEEEbiography}[{\includegraphics[width=1in,height=1.25in,clip,keepaspectratio]{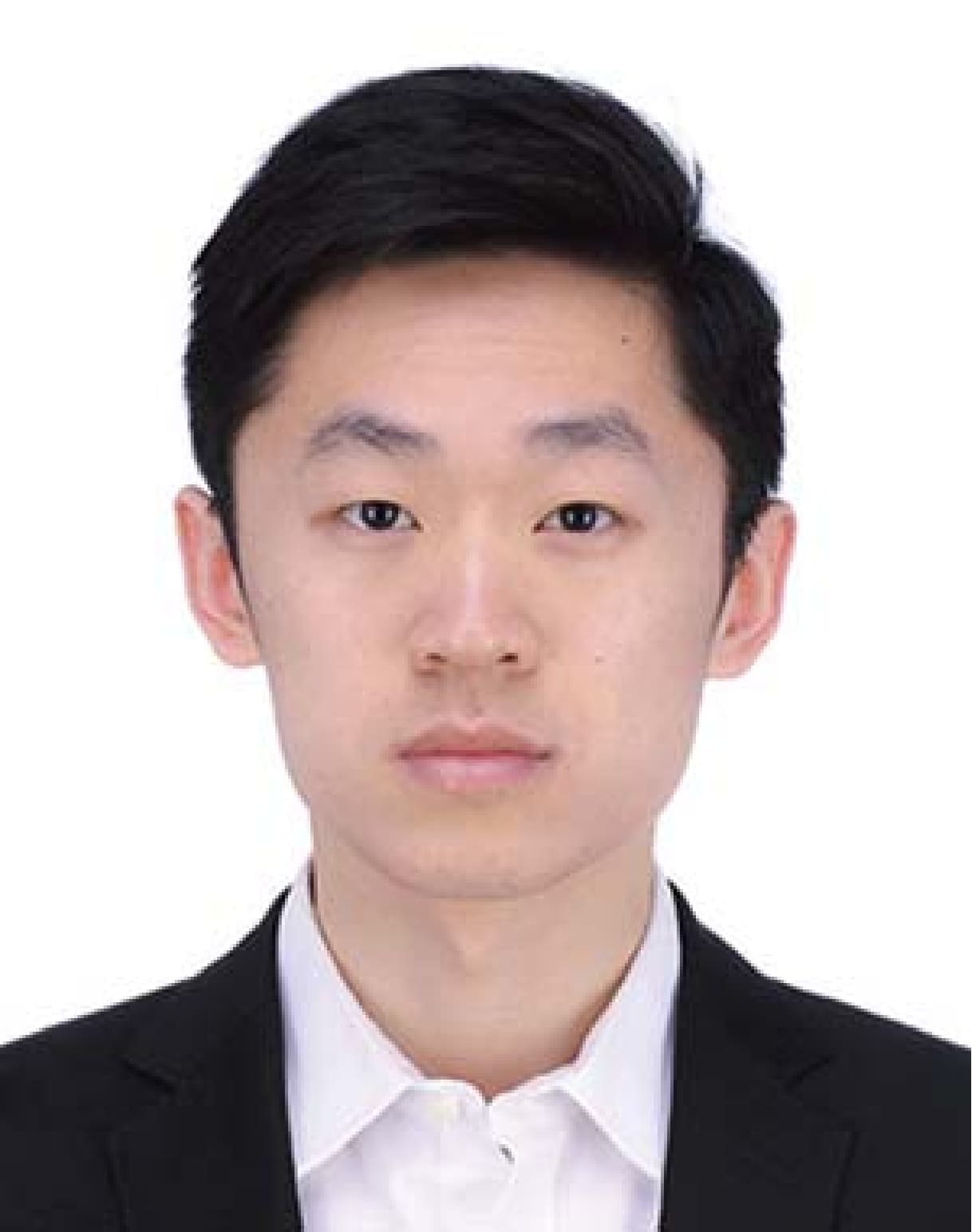}}]{Jingtao Ding}
received the BS degrees in electronic engineering and the PhD degree in electronic engineering from Tsinghua University, Beijing, China, in 2015 and 2020, respectively. He is currently a postdoctoral research fellow with the Department of Electronic Engineering, Tsinghua University. His research interests include mobile computing, spatiotemporal data mining and user behavior modeling. He has more than 60 publications in journals and conferences such as IEEE Transactions on Knowledge and Data Engineering, ACM Transactions on Information Systems, KDD, NeurIPS, WWW, ICLR, SIGIR, IJCAI, etc.
\end{IEEEbiography}

\vspace*{-1.1cm}
\begin{IEEEbiography}[{\includegraphics[width=1in,height=1.25in,clip,keepaspectratio]{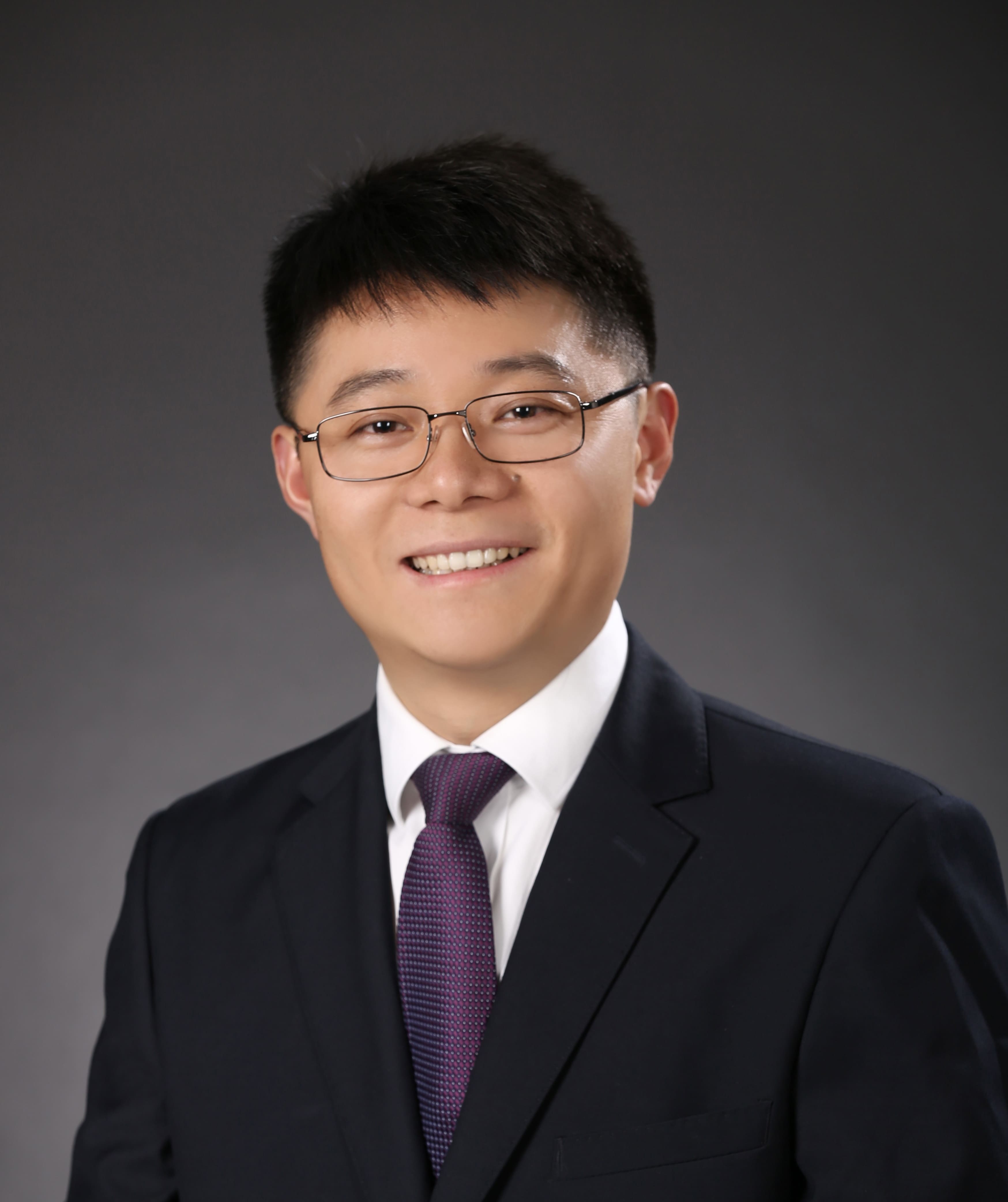}}]{Yong Li}
(Senior Member, IEEE) received the PhD degree in electronic engineering from Tsinghua University in 2012. He is currently a tenured professor with the Department of Electronic Engineering, Tsinghua University. He has authored or coauthored more than 100 papers on first-tier international conferences and journals, including KDD, WWW, UbiComp, SIGIR, AAAI, IEEE Transactions on Knowledge and Data Engineering, IEEE Transactions on Mobile Computing and his papers have total citations more than 36,000. His research interests include machine learning and big data mining, particularly, automatic machine learning and spatial-temporal data mining for urban computing, recommender systems, and knowledge graphs. He was the general chair, the TPC chair, an SPC or a TPC member for several international workshops and conferences. He is on the editorial board of two IEEE journals. Among them, 10 are ESI Highly Cited Papers in computer science, and five were the recipient of conference Best Paper (run-up) awards.
\end{IEEEbiography}

\clearpage
\appendix

\subsection{Equivalence Between Predicting Edge Generation Times and Pairwise Orderings}\label{apdx:equal}

This section formalizes the connection between regressing edge generation times and predicting pairwise temporal orderings. Consider a network with \(M\) edges generated sequentially. Let \(\alpha_i\in[0,1]\) denote the normalized generation time of edge \(e_i\), where smaller values indicate earlier formation. Without loss of generality, we index edges in chronological order so that \(\alpha_i=i/M\).

Let \(p\) denote the accuracy of a pairwise ordering predictor, consistent with Section~\ref{sec:pre}: for a randomly sampled edge pair \((e_i,e_j)\), it correctly predicts which edge forms later with probability \(p\). Define the pairwise outcome
\[
Y_{ij} =
\begin{cases}
1, & \text{predict } \alpha_i > \alpha_j,\\
0, & \text{predict } \alpha_i < \alpha_j,
\end{cases}
\]
so that
\[
\mathbb{E}(Y_{ij}) =
\begin{cases}
p, & \alpha_i > \alpha_j,\\
1-p, & \alpha_i < \alpha_j,
\end{cases}
\qquad
\mathrm{Var}(Y_{ij}) = p(1-p).
\]

We aggregate pairwise outcomes using a Borda-style score
\begin{equation}
	u_i = \frac{1}{M-1}\sum_{j\ne i} Y_{ij}.
	\label{eq:apdx_borda_score}
\end{equation}
Since edge \(i\) is later than exactly \(i-1\) edges and earlier than \(M-i\) edges, we obtain
\[
\mathbb{E}(u_i)
=
\frac{(i-1)p + (M-i)(1-p)}{M-1},
\]
which is linear in \(i\). Therefore, the order induced by \(\{u_i\}\) recovers the chronological edge order in expectation.

Moreover, since \(u_i\) averages \(M-1\) Bernoulli variables, its variance satisfies
\[
\mathrm{Var}(u_i)
=
\frac{p(1-p)}{M-1}
\approx
\frac{p(1-p)}{M}.
\]
Because \(\mathbb{E}(u_i)\) varies with slope proportional to \(2p-1\), the uncertainty in the recovered normalized position scales as
\[
\mathcal{E}_\text{theory}
\propto
\sqrt{\frac{p(1-p)}{(2p-1)^2}}
\frac{1}{\sqrt{M}},
\]
which matches Eq.~\eqref{eq:equivalence} in the main text up to the negligible \(M\) vs.\ \(M-1\) normalization. This result shows that (i) the global ordering error decreases as \(M\) increases and (ii) a moderately accurate pairwise predictor (\(p>0.5\)) can already support reliable global sequence recovery, justifying pairwise learning as a stable surrogate for edge-time regression.

\subsection{High-Order Structure in Steady States}\label{apdx:steady_highorder}

This section provides theoretical intuition for why steady states can encode high-order structural information beyond local topology. In widely used network dynamics, including diffusion processes, opinion dynamics, and linearized gene-regulation models, state evolution can often be written as linear discrete- or continuous-time dynamics governed by random-walk operators or graph Laplacians, e.g.,
\begin{equation}
	x(t+1) = P x(t), \qquad P = D^{-1}A,
\end{equation}
or in continuous time,
\begin{equation}
	\frac{dx}{dt} = -\mathcal{L} x(t),
\end{equation}
where $P$ is the random-walk matrix and $\mathcal{L}$ is the graph Laplacian. Under appropriate stability conditions, the system converges to a steady state $x^*$, which can be expressed as
\begin{equation}
	x^* = \lim_{t\to\infty} P^t x(0)
	= \sum_{k=0}^{\infty} P^k x(0).
\end{equation}
This expression highlights a path-accumulation view: the steady state is not determined by a single adjacency interaction, but integrates contributions from multi-hop propagation paths through higher powers of $P$. Consequently, high-order topological patterns---such as inter-community bridges, bottlenecks, and global centrality heterogeneity---can accumulate in $x^*$ even when they are not evident from local degrees or one-hop neighborhoods. From this viewpoint, $x^*$ is an integrated representation of multi-hop connectivity rather than a simple smoothing of local topology.

A complementary explanation follows from spectral analysis. By eigen-decomposing the diffusion operator, we have
\begin{equation}
	P = \sum_{i} \lambda_i u_i u_i^\top, \qquad
	\mathcal{L} = \sum_{i} \mu_i v_i v_i^\top,
\end{equation}
and the steady state can be written as
\begin{equation}
	x^* = \sum_{i} f(\lambda_i)\,(u_i^\top x(0))\,u_i,
\end{equation}
where $u_i$ and $\lambda_i$ are eigenvectors and eigenvalues, and $f(\lambda_i)$ is a decay function determined by the specific dynamics. Eigenvectors encode global structural modes (e.g., community-separation directions and centrality gradients), while eigenvalues control how strongly each mode persists in long-term dynamics. As a result, high-order properties such as community organization, bridge structure, and global diffusivity can shape the spatial distribution of $x^*$ through the spectrum. In other words, $x^*$ can be viewed as a filtered spectral representation of global topology beyond local degrees or one-hop adjacency.

In summary, steady states admit two complementary interpretations: cumulative multi-hop propagation and weighted superposition of spectral modes. Both indicate that $x^*$ carries information well beyond local topology, including multi-hop paths, community structure, and global connectivity patterns, providing theoretical support for using steady-state observations in edge-order recovery.

\subsection{Path Dependence of Steady States}\label{apdx:steady_pathdependence}

This section explains why steady states can be informative of edge formation order through \emph{path dependence}. When network structure and node dynamics coevolve, each fixed topology $G$ corresponds to a steady state $x^*(G)$. If the structure remains unchanged for a period, node states converge toward $x^*(G)$. However, when the network evolves through a sequence of edge formation events, the topology changes over time as
\[
G^{(0)},\,G^{(1)},\,\dots,\,G^{(T)},
\]
the system state typically cannot fully reach the corresponding instantaneous steady state at each stage; instead, it approaches it at a finite rate. Because the trajectory cannot track the moving steady-state branch pointwise, the final state becomes an accumulated response to the entire path of instantaneous equilibria, and is thus sensitive to the \emph{path} of structural changes. Therefore, even if two networks share the same final topology, different edge formation orders can generally lead to different observed steady states.

To formalize this phenomenon, we consider a generic relaxation dynamics:
\begin{equation}
	\dot{x}(t) = -\bigl(x(t)-x^*(G(t))\bigr),
	\label{eq:relax_general}
\end{equation}
where $x(t)\in\mathbb{R}^n$ denotes the system state and $x^*(G(t))$ is the instantaneous steady state associated with the current topology $G(t)$. This abstraction only assumes the generic property that, under a fixed structure, the dynamics relaxes toward a steady state, without imposing model-specific details.

For any given structural evolution path $G:[0,T]\to\mathcal{G}$, the solution to~\eqref{eq:relax_general} can be written explicitly as
\begin{equation}
	x(T) = e^{-T} x(0) + \int_{0}^{T} e^{-(T-s)}\, x^*(G(s))\,ds.
	\label{eq:solution_general}
\end{equation}
Eq.~\eqref{eq:solution_general} shows that the final state consists of an exponentially decayed initial condition and a time-weighted integral over the instantaneous steady states $x^*(G(s))$. The integral term implies that the observed final state is not determined solely by the terminal structure $G^{(T)}$, but depends on the entire trajectory of instantaneous equilibria along $s\in[0,T]$.

Therefore, if two structural evolution paths $G_A(t)$ and $G_B(t)$ differ on $[0,T]$ while sharing the same final structure $G_A(T)=G_B(T)$, then
\[
\int_0^T e^{-(T-s)} x^*(G_A(s))\,ds 
\;\neq\;
\int_0^T e^{-(T-s)} x^*(G_B(s))\,ds,
\]
and Eq.~\eqref{eq:solution_general} directly implies
\[
x_A(T) \neq x_B(T).
\]
In other words, under structure--dynamics coevolution, the observed final state depends on the \emph{evolution path} of topology rather than only the terminal snapshot. This provides a mechanism by which steady-state observations can carry discriminative information about edge formation order.

This mechanism is mathematically analogous to rate-induced tipping (R-tipping). In the R-tipping framework, a system has a time-varying parameter $\lambda(t)$ and an associated branch of instantaneous equilibria $x^*(\lambda)$; when the parameter changes at a rate comparable to the system's intrinsic relaxation, trajectories can fail to track the equilibrium branch, becoming sensitive to the parameter path rather than the terminal value. In coevolving network dynamics, the topology $G(t)$ plays the role of a time-dependent parameter and $x^*(G(t))$ forms the instantaneous steady-state branch. Because the system has limited tracking capacity during structural changes, the final state becomes a time-weighted response to the entire branch $x^*(G(t))$, and different edge formation orders lead to different instantaneous steady-state trajectories and hence observable steady-state differences.

\subsection{Global Ordering Consistency via Borda Aggregation}\label{apdx:borda_spearman}

This section reports global ordering consistency under \textit{Gene} and \textit{Opinion} dynamics. We reconstruct a complete edge sequence by aggregating pairwise precedence predictions using Borda count, and measure its agreement with the ground truth using Spearman's rank correlation $\rho$ (Table~\ref{tab:spearman_rho_gene_opinion}).
Overall, \textit{CS$^2$} consistently achieves the best global consistency. Under \textit{Gene} dynamics, it attains Avg $\rho=\textit{0.8179}$, improving over the strongest baseline by 3.8\%--13.6\% across datasets. Under \textit{Opinion} dynamics, it achieves Avg $\rho=\textit{0.8109}$ with improvements of 1.8\%--11.6\%. These results confirm that the gains of CS$^2$ in global order recovery are robust across different dynamical processes.

\begin{table*}[t]
	\centering
	\caption{Global ordering consistency (Spearman-$\rho$) under \textit{Gene} and \textit{Opinion} dynamics. Higher is better. The rightmost column (Avg) reports the mean $\rho$ over all datasets, and the best baseline results are \underline{underlined}.}
	\label{tab:spearman_rho_gene_opinion}
	\renewcommand{\arraystretch}{1.2}
	\setlength{\tabcolsep}{6pt}
	
	\begin{tabular}{llcccccc>{\columncolor[gray]{0.9}}c}
		\toprule
		\textbf{Dynamics} & \textbf{Method} & \textbf{Chaos} & \textbf{Complex} & \textbf{Fluctuations} & \textbf{Interfaces} & \textbf{Phase} & \textbf{Thermo} & \textbf{Avg} \\
		\midrule
			\multirow{12}{*}{\textit{Gene}}
			& StateEvo      & 0.1807 & 0.2454 & 0.1445 & 0.1686 & 0.0978 & 0.4530 & 0.2150 \\
			& StructEvo     & 0.2831 & 0.2749 & 0.2786 & 0.3489 & 0.2909 & 0.4016 & 0.3130 \\
			& StructStaEvo  & 0.2719 & 0.1428 & 0.2533 & 0.3592 & 0.2889 & 0.2409 & 0.2595 \\
			& TopoDiff      & 0.2012 & 0.1216 & 0.1764 & 0.1723 & 0.2149 & 0.4582 & 0.2241 \\
			& TopoDiffStaEvo & 0.1570 & 0.2166 & 0.1804 & 0.2175 & 0.1947 & 0.3846 & 0.2251 \\
			& N2VecEvo      & 0.7309 & \underline{0.7885} & 0.7727 & 0.8234 & 0.6176 & 0.7008 & 0.7390 \\
			& N2VecStaEvo   & \underline{0.7367} & 0.7411 & \underline{0.7955} & \underline{0.8420} & 0.6802 & 0.6234 & 0.7365 \\
			& LLaCEEvo      & 0.6687 & 0.7681 & 0.7136 & 0.8182 & 0.7071 & 0.6757 & 0.7252 \\
			& LLaCEStaEvo   & 0.6460 & 0.7467 & 0.7694 & 0.8282 & \underline{0.7098} & \underline{0.7084} & 0.7348 \\
			& TEREvo        & 0.6183 & 0.7699 & 0.6584 & 0.7394 & 0.6266 & 0.4612 & 0.6456 \\
			& TERStaEvo     & 0.5574 & 0.6871 & 0.6356 & 0.7541 & 0.6041 & 0.6512 & 0.6482 \\
			& \textbf{CS$^2$}
		& \makecell{\textbf{0.7757} \\ (+5.3\%)}
		& \makecell{\textbf{0.8387} \\ (+6.4\%)}
		& \makecell{\textbf{0.8258} \\ (+3.8\%)}
		& \makecell{\textbf{0.8923} \\ (+6.0\%)}
		& \makecell{\textbf{0.8066} \\ (+13.6\%)}
		& \makecell{\textbf{0.7685} \\ (+8.5\%)}
		& \textbf{0.8179} \\
		\midrule
			\multirow{12}{*}{\textit{Opinion}}
			& StateEvo      & 0.1607 & 0.3617 & 0.1175 & 0.1094 & 0.2343 & 0.3858 & 0.2282 \\
			& StructEvo     & 0.2831 & 0.2749 & 0.2786 & 0.3489 & 0.2909 & 0.4016 & 0.3130 \\
			& StructStaEvo  & 0.3169 & 0.1585 & 0.4116 & 0.3793 & 0.4591 & 0.4415 & 0.3612 \\
			& TopoDiff      & 0.2012 & 0.1216 & 0.1764 & 0.1723 & 0.2149 & 0.4582 & 0.2241 \\
			& TopoDiffStaEvo & 0.1806 & 0.2920 & 0.1264 & 0.1801 & 0.2193 & 0.4439 & 0.2404 \\
			& N2VecEvo      & 0.7309 & \underline{0.7885} & 0.7727 & 0.8234 & 0.6176 & 0.7008 & 0.7390 \\
			& N2VecStaEvo   & \underline{0.7367} & 0.7411 & \underline{0.7955} & \underline{0.8420} & 0.6802 & 0.6234 & 0.7365 \\
			& LLaCEEvo      & 0.6687 & 0.7681 & 0.7136 & 0.8182 & 0.7071 & 0.6757 & 0.7252 \\
			& LLaCEStaEvo   & 0.6460 & 0.7467 & 0.7694 & 0.8282 & \underline{0.7098} & \underline{0.7084} & 0.7348 \\
			& TEREvo        & 0.6183 & 0.7699 & 0.6584 & 0.7394 & 0.6266 & 0.4612 & 0.6456 \\
			& TERStaEvo     & 0.5574 & 0.6871 & 0.6356 & 0.7541 & 0.6041 & 0.6512 & 0.6482 \\
			& \textbf{CS$^2$}
		& \makecell{\textbf{0.7590} \\ (+3.0\%)}
		& \makecell{\textbf{0.8479} \\ (+7.5\%)}
		& \makecell{\textbf{0.8100} \\ (+1.8\%)}
		& \makecell{\textbf{0.8850} \\ (+5.1\%)}
		& \makecell{\textbf{0.7727} \\ (+8.9\%)}
		& \makecell{\textbf{0.7905} \\ (+11.6\%)}
		& \textbf{0.8109} \\
		\bottomrule
	\end{tabular}
\end{table*}

\subsection{Global Ordering Consistency under Gene and Opinion Dynamics}\label{apdx:rank_scatter_gene_opinion}

To complement the Spearman-$\rho$ results in Table~\ref{tab:spearman_rho_gene_opinion}, we visualize global ordering consistency under \textit{Gene} and \textit{Opinion} dynamics using the same binned protocol as the SIS case in the main text. Specifically, we partition edges into 10 percentile bins by their ground-truth ranks and plot, for each bin, the median predicted rank with an intra-bin variance band.
We further quantify the deviation from the ideal $y=x$ line by the RMSE between the binned median curve and the diagonal reference (reported in the legend and in each panel). CS$^2$ achieves the lowest RMSE across all datasets, with average RMSE of 0.059 under \textit{Gene} dynamics and 0.068 under \textit{Opinion} dynamics, substantially improving over TERStaEvo (0.111) and StructEvo (0.207). These results indicate that CS$^2$ yields more accurate and stable global recovery across evolution stages, consistent with its higher Spearman-$\rho$ in Table~\ref{tab:spearman_rho_gene_opinion}.

\begin{figure*}[t]
	\centering
\includegraphics[width=0.92\textwidth]{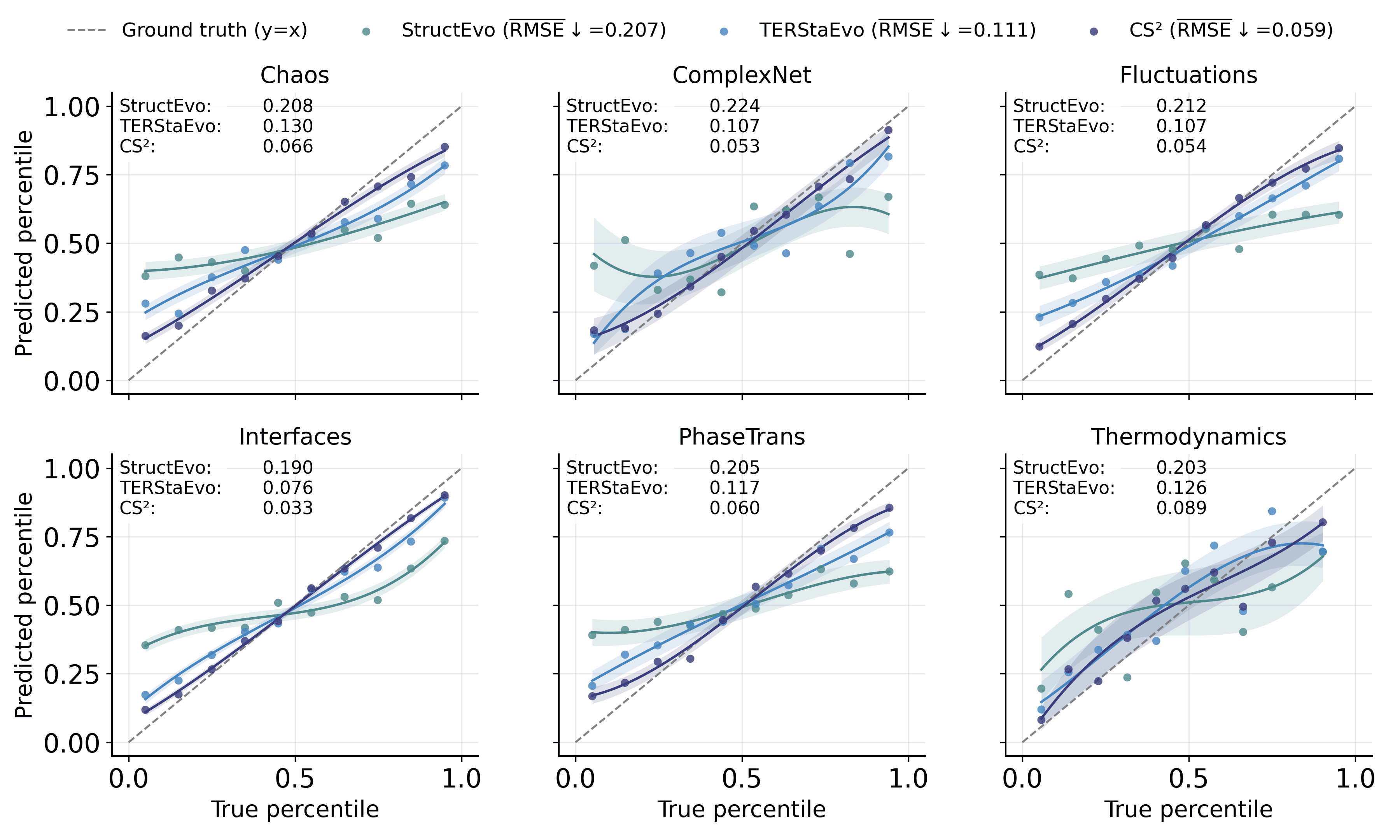}
	\caption{Binned global ordering consistency under \textit{Gene} dynamics across datasets.}
	\label{fig:rank_scatter_gene}
\end{figure*}

\begin{figure*}[t]
	\centering
\includegraphics[width=0.92\textwidth]{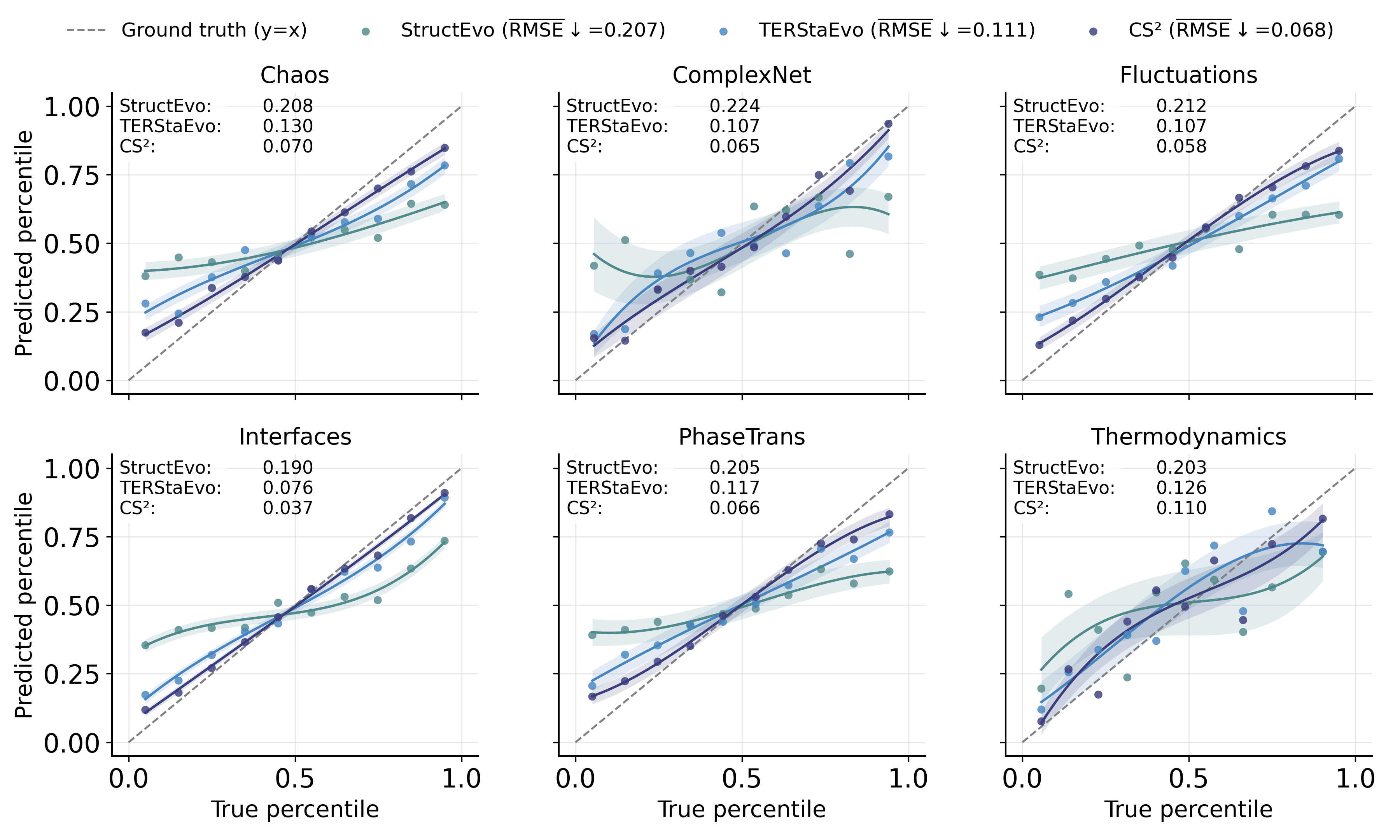}
	\caption{Binned global ordering consistency under \textit{Opinion} dynamics across datasets.}
	\label{fig:rank_scatter_opinion}
\end{figure*}

\subsection{Global Property Recovery under Gene and Opinion Dynamics}\label{apdx:global_property_gene_opinion}

\begin{figure*}[t]
	\centering
\includegraphics[width=0.99\textwidth]{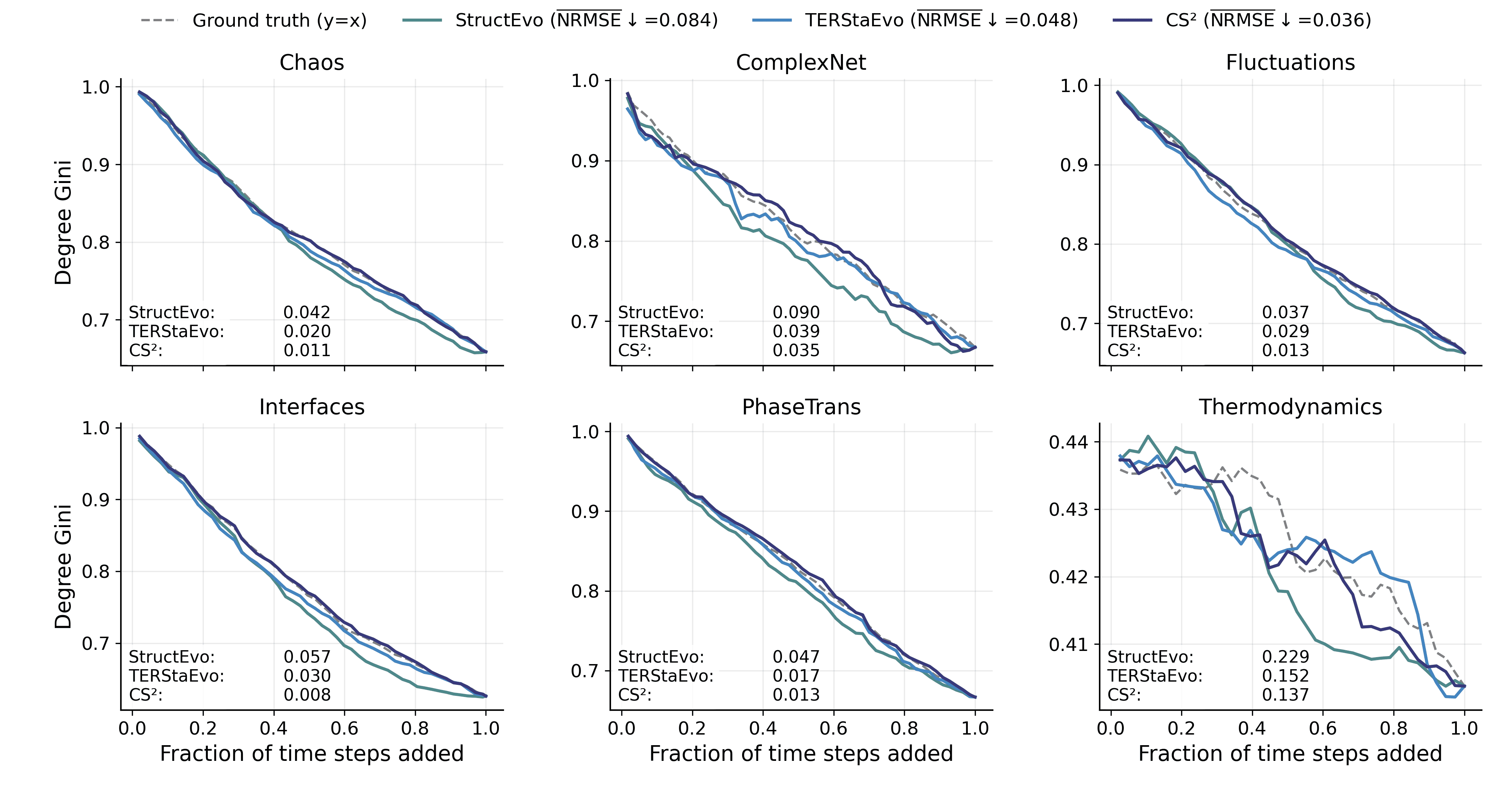}
	\caption{Evolution of degree Gini coefficient across datasets under \textit{Gene} dynamics.}
	\label{fig:prop_gene_degree_gini}
\end{figure*}

\begin{figure*}[t]
	\centering
\includegraphics[width=0.99\textwidth]{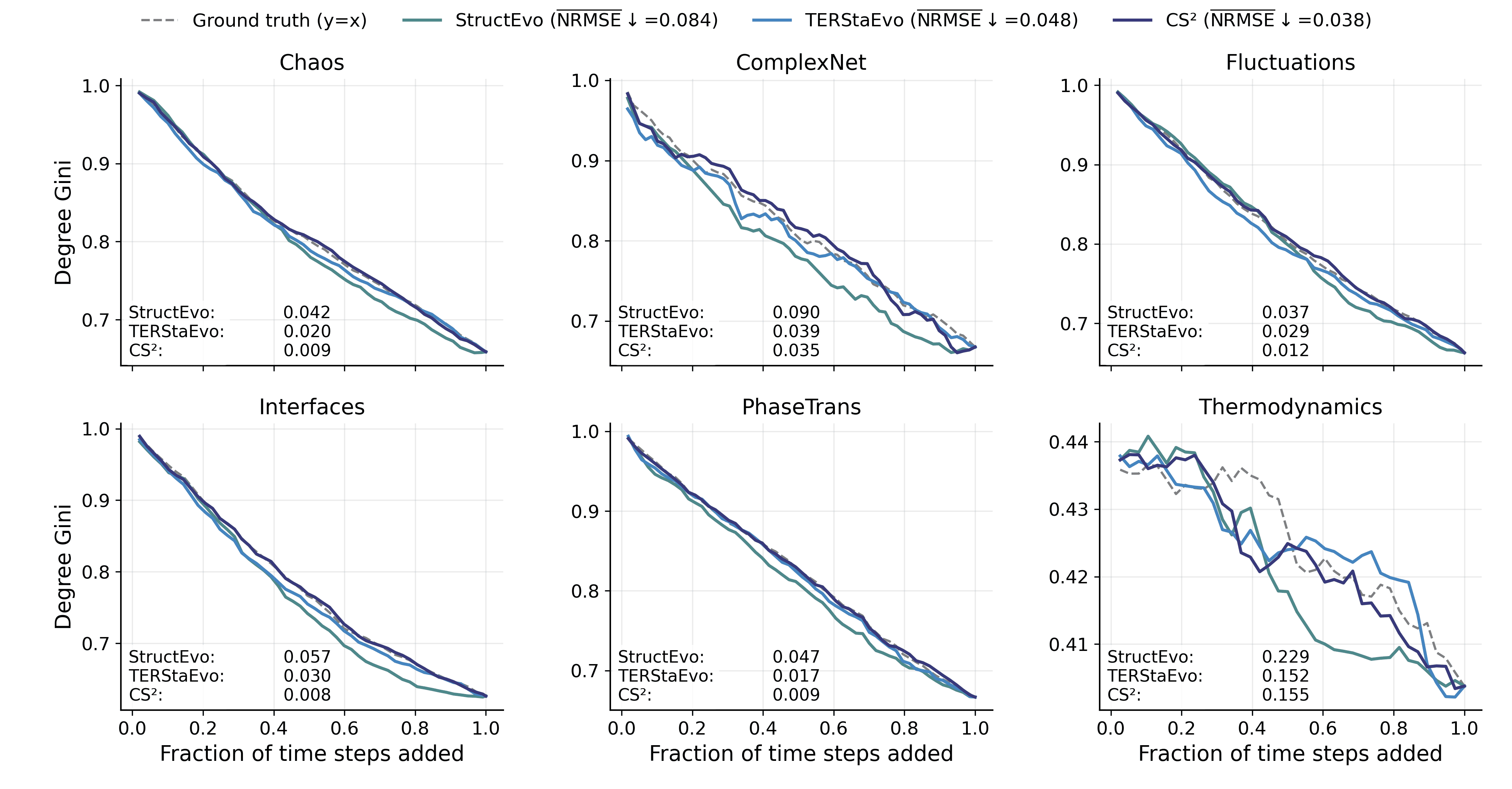}
	\caption{Evolution of degree Gini coefficient across datasets under \textit{Opinion} dynamics.}
	\label{fig:prop_opinion_degree_gini}
\end{figure*}

\begin{figure*}[t]
	\centering
\includegraphics[width=0.99\textwidth]{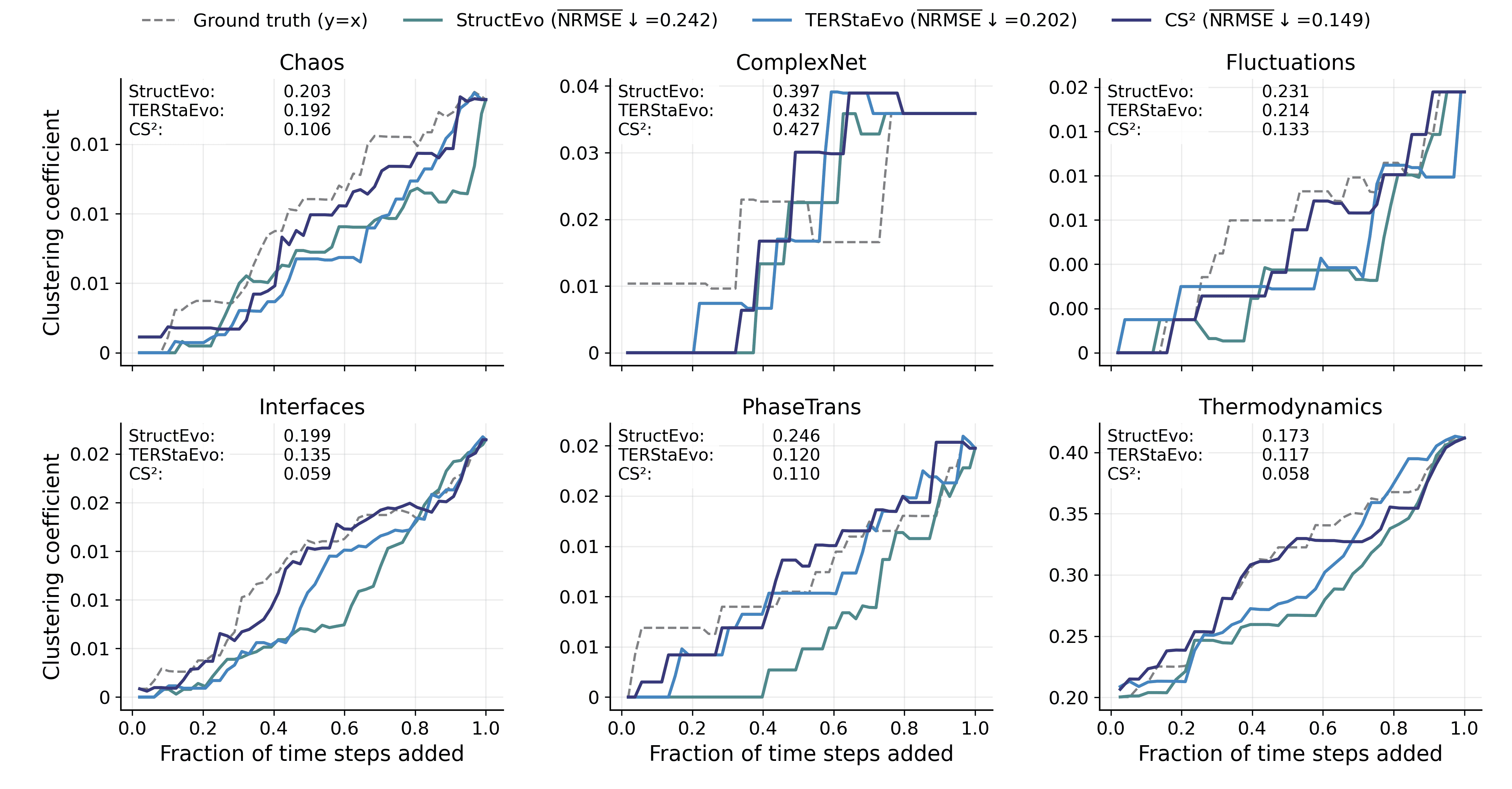}
	\caption{Evolution of average clustering coefficient across datasets under \textit{Gene} dynamics.}
	\label{fig:prop_gene_clustering}
\end{figure*}

\begin{figure*}[t]
	\centering
\includegraphics[width=0.99\textwidth]{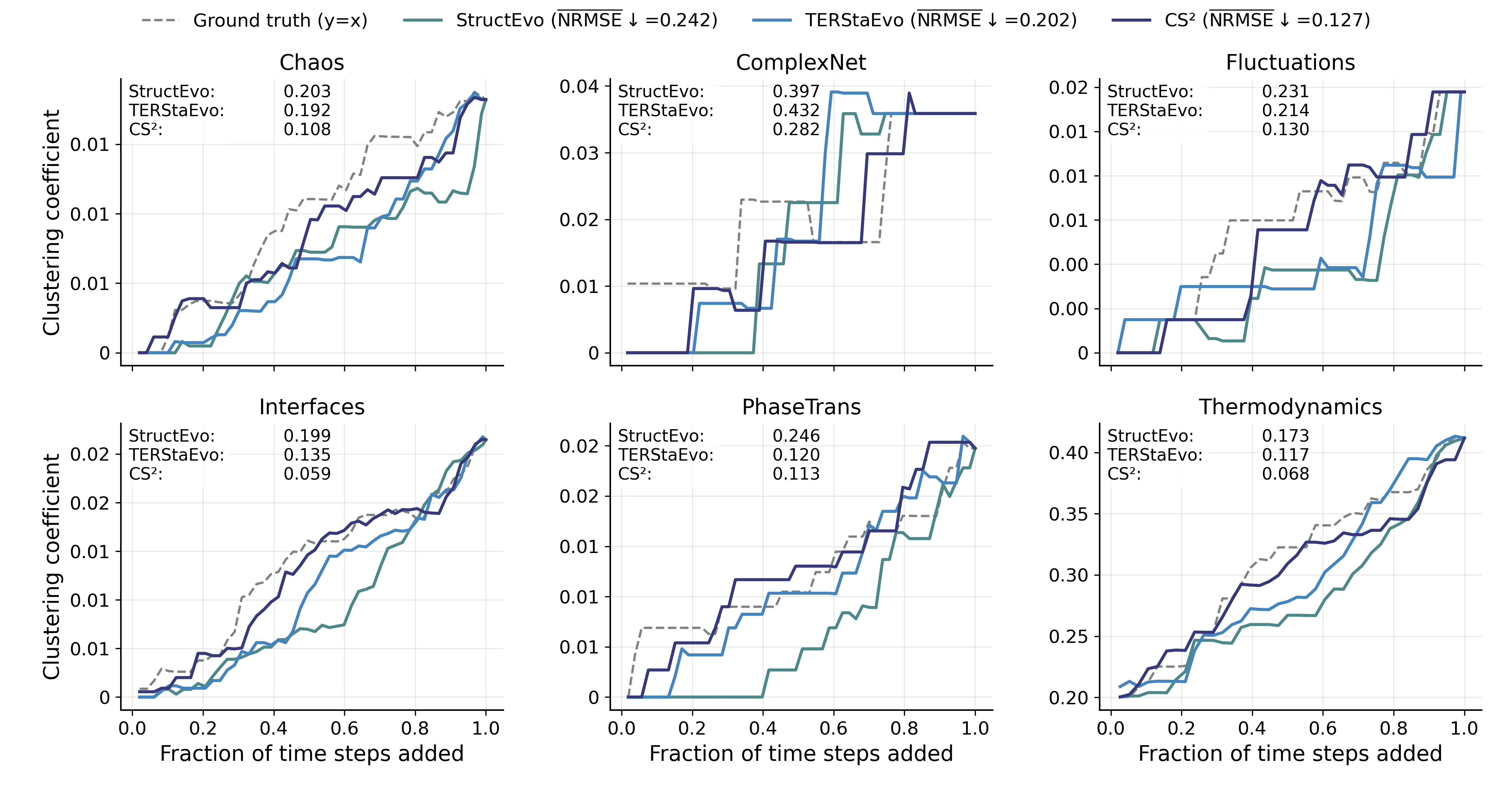}
	\caption{Evolution of average clustering coefficient across datasets under \textit{Opinion} dynamics.}
	\label{fig:prop_opinion_clustering}
\end{figure*}

\begin{figure*}[t]
	\centering
\includegraphics[width=0.99\textwidth]{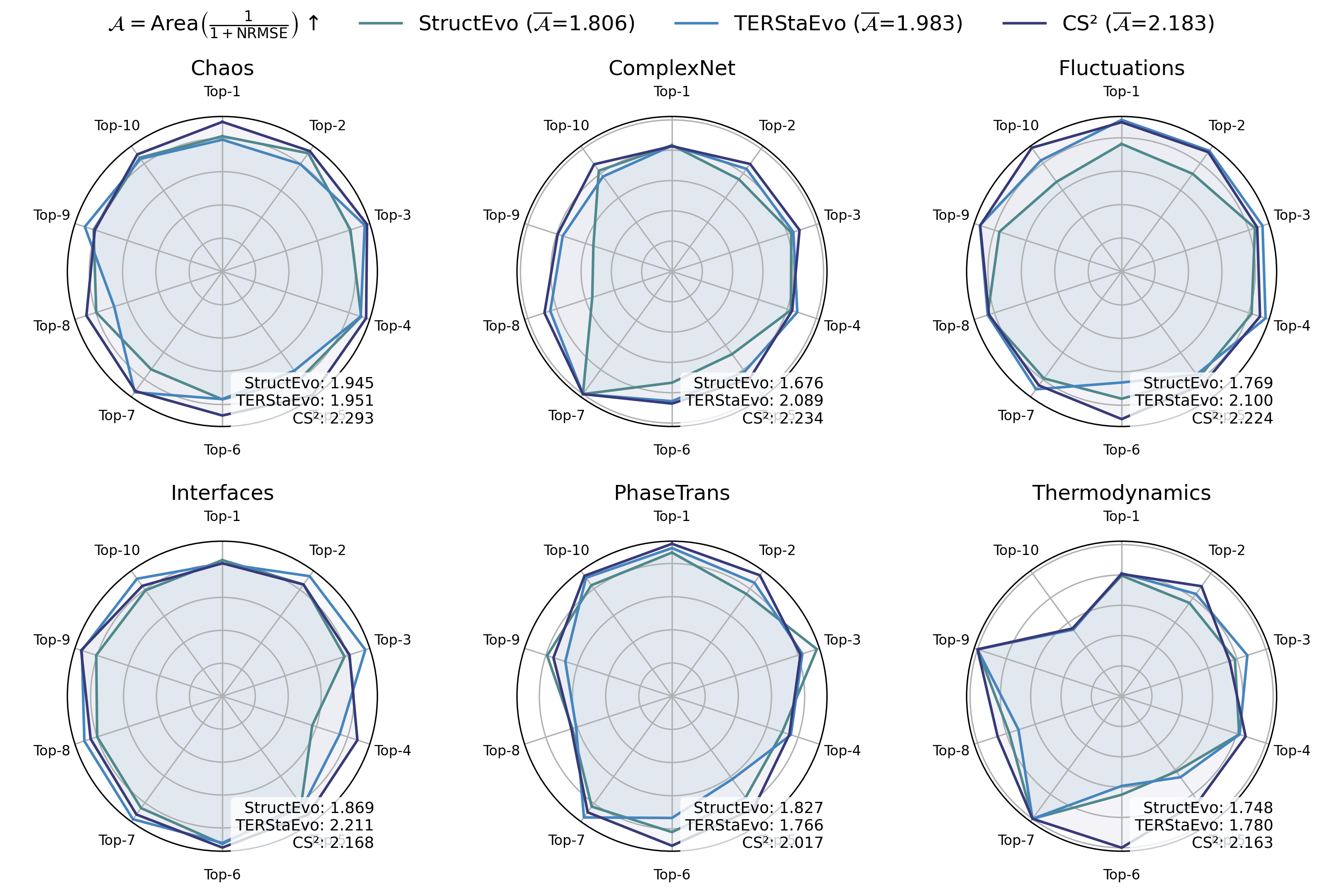}
	\caption{Top-hub degree evolution fidelity across datasets under \textit{Gene} dynamics.}
	\label{fig:prop_gene_hubs}
\end{figure*}

\begin{figure*}[t]
	\centering
\includegraphics[width=0.99\textwidth]{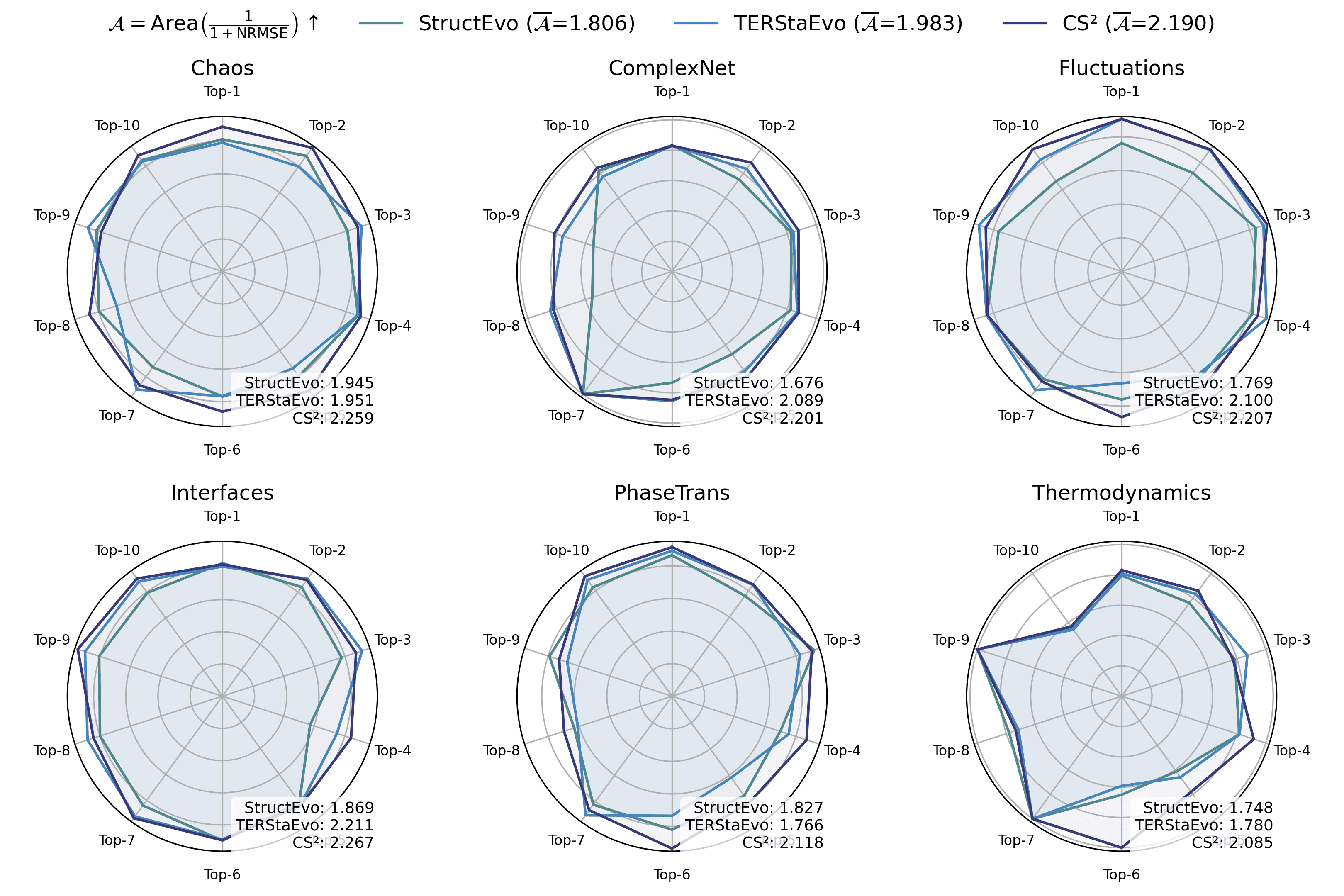}
	\caption{Top-hub degree evolution fidelity across datasets under \textit{Opinion} dynamics.}
	\label{fig:prop_opinion_hubs}
\end{figure*}

Figures~\ref{fig:prop_gene_degree_gini}--\ref{fig:prop_opinion_hubs} report global property recovery under \textit{Gene} and \textit{Opinion} dynamics. Across both dynamics, CS$^2$ provides the most faithful reconstruction of macroscopic trajectories. In particular, it achieves the lowest errors for \textit{Degree Gini} (NRMSE$\downarrow$: 0.036 for \textit{Gene} and 0.038 for \textit{Opinion}), improving over the best baseline (TERStaEvo: 0.048) by 25.0\% and 20.8\%, respectively. For \textit{Clustering}, CS$^2$ also yields the lowest NRMSE (0.149 and 0.127), improving over TERStaEvo (0.202) by 26.2\% and 37.1\%. For \textit{Hub radar}, CS$^2$ attains the highest area-based fidelity score ($\overline{\mathcal{A}}{=}2.183$ and $2.190$), exceeding TERStaEvo ($\overline{\mathcal{A}}{=}1.983$) by 10.1\% and 10.4\%. These results indicate that the improved ordering consistency of CS$^2$ translates into more accurate recovery of degree-heterogeneity formation, triangle closure, and hub growth.

\end{document}